\documentclass[reprint,aps,prd,nofootinbib,a4paper,10pt,twocolumn,superscriptaddress]{revtex4-2}

\usepackage{hyperref}
\usepackage{amssymb,amsmath,latexsym,mathrsfs}
\usepackage{graphicx,subfigure}
\usepackage{epsfig}
\usepackage{varioref,xr-hyper}
\usepackage{color}
\usepackage{multirow}
\usepackage{array}
\hypersetup{
	colorlinks   = true,
	citecolor    = blue
}

\begin{document} 
	
	\title{Redshift drift in a universe with structure II: Light rays propagated through a Newtonian N-body simulation}
	
	\author{Sofie Marie Koksbang} 
	\email{koksbang@cp3.sdu.dk}
	\affiliation{CP$^3$-Origins, University of Southern Denmark, Campusvej 55, DK-5230 Odense M, Denmark}

\begin{abstract} 
\noindent
The redshift drift is computed along light rays propagating through a simulated universe based on the Newtonian N-body simulation code GADGET-2 combined with a perturbed Friedmann-Lemaitre-Robertson-Walker metric in the Newtonian gauge. It is found that the mean redshift drift is equal to the drift of the mean redshift to the precision of the numerical computations and that this is due to a high degree of cancellation between two dominant components of the redshift drift. This result is contrary to earlier findings based on inhomogeneous cosmological models exhibiting cosmic backreaction.
\newline\indent
For simplicity, the results neglect contributions from optical drift. Based on a study of the redshift drift in a Lemaitre-Tolman-Bondi model, the optical drift effects are estimated to be at most of order 10\% of the redshift drift signal. In addition, it is found that the redshift drift contribution from peculiar acceleration of the emitter is negligible in the simulation setup. However, it is expected that the contribution from peculiar acceleration of the emitter is suppressed in the setup due to low resolution of structures and it is hence expected that this contribution will be larger for real observations.
\end{abstract}
	
	\maketitle
	
\section{Introduction}
Obtaining new observational data is key for understanding the reason(s) behind the discrepancies between the standard $\Lambda$CDM model and various observational data sets, discussed in e.g. \cite{tensions1, tensions2, tensions3, tensions4, tensions5, tensions6, tensions7, tension_added_sunny_1,tension_added_sunny_2}. Such new data should not only include more of the same types of data that we already have, but also new types of observations. A prime example of a ``new'' type of observable is the redshift drift, i.e. the change of a source's redshift in time.
\newline\indent
Redshift drift measurements were envisioned already in 1962 in \cite{sandage,mcvittie}. In the former of these, it was concluded that it would require at the order of $10^7$ years of observation time for the signal to be large enough for detection. With today's technology, it is instead estimated that the measurements may be feasible within only a few decades \cite{bolejko_flux,SKA, feasible1, feasible2, feasible3, feasible4, feasible5}, although it will not be an easy achievement.
\newline\indent
The possibility of actually measuring the redshift drift within our lifetime has spurred a significant amount of research into redshift drift in recent years. The redshift drift has for instance been studied in various homogeneous and isotropic cosmological models such as in \cite{dz_hom1, dz_hom2, dz_hom3} but also in less standard scenarios with e.g. modified gravity theories \cite{dz_FR} and with a varying speed of light \cite{dz_varying_c}. In addition, the redshift drift has been considered in a variety of different inhomogeneous and/or anisotropic models such as in Lemaitre-Tolman-Bondi \cite{dz_LTB1,dz_LTB2,dz_LTB3, dz_LTB4}, Szekeres \cite{dz_SZ1,dz_Sz2}, Stephani \cite{dz_stephani1,dz_stephani2}, Bianchi I \cite{dz_BianchiI_1, dz_BianchiI_2,dz_pert_bianci}, perturbative, \cite{dz_pert_bianci, Linder_dz1, Linder_dz2} and Einstein-Strauss models \cite{dz_ES}, and has been utilized to develop different cosmological tests \cite{dz_test1,dz_test2,dz_test3} (see e.g. \cite{dz_LTB4} regarding a correction of the first of these).
\newline\indent
The redshift drift is a particularly interesting observable because it within the Friedmann-Lemaitre-Robertson-Walker (FLRW) cosmologies is given by
\begin{align}\label{eq:dz_FLRW}
\begin{split}
    \delta z &= \delta t_0\left[(1+z)H_0-H(z) \right] \\
    &= (1+z)\left[a_{,t}(t_0) - a_{,t}(t_e)  \right] ,
\end{split}
\end{align}
where $\delta t_0$ is the time interval and $t_e$ is the emission time. Clearly, if $H_0$ is known by other means, the redshift drift yields a direct measurement of the expansion rate of the FLRW universe. A direct measurement of $H(z)$ is important in its own right since today's observables only indirectly teach us about the expansion rate of the Universe. However, the importance of measuring the redshift drift becomes even more clear when noticing that the sign of the redshift drift can become positive only if the expansion rate has accelerating periods. This is clear for the FLRW models as seen by equation \ref{eq:dz_FLRW} but an important follow-up question is to what extent this extrapolates to other cosmological models. It was in \cite{another_look} shown that the redshift drift in a model universe with average accelerated expansion generated by cosmic backreaction \cite{fluid1,fluid2,fluid3} was negative. In other words, it was shown that the observed redshift drift in the model was negative despite structures leading to apparent accelerated expansion and based on that it was in \cite{another_look} conjectured that the redshift drift in a statistically homogeneous and isotropic universe will be positive only if the Universe is undergoing local accelerated expansion due to dark energy -- a conjecture that has later been supported by e.g. \cite{Asta_dz_DE}. On the other hand, within the ambitious and quite complex inhomogeneous cosmological model known as timescape cosmology, the redshift drift can still be positive at low redshift \cite{timescape,timescape2} without local accelerated expansion induced by dark energy.
\newline\newline
With the expectation that redshift drift measurements are about to become feasible it is necessary to understand possible sources of measurement errors that need to be taken into account when determining the precision of the measurements. One important question in this regard is how big the structure-induced fluctuations of redshift drift signals are, assuming a standard cosmological scenario where structures are well described by e.g. Newtonian N-body simulations. The objective with the work presented here is therefore to compute the redshift drift along light rays in such a simulation to quantify the fluctuations in the measurements around the values expected based on equation \ref{eq:dz_FLRW}. This complements earlier studies such as \cite{feasible1, pec_acc_not1, pec_acc_not2, pec_acc_not3} which indicate that the peculiar velocities of sources have only a small impact on the observed redshift drift signal.
\newline\indent
The examples of \cite{another_look,dz_BianchiI_2, Asta_first_dz} show that the drift of the mean redshift will not in general be equal to the mean drift of the redshift. On the other hand, the results of \cite{dz_LTB2,dz_LTB3} show that the mean redshift drift {\em does} follow the drift of the mean redshift in LTB Swiss cheese models. These results hint towards an important question, namely whether or not we can expect the mean redshift drift to equal the drift of the mean redshift in our universe. While the answer to this question seems to e.g. depend on whether or not there is significant cosmic backreaction in the real universe, the standard cosmological description of the inhomogeneous universe is based on Newtonian N-body simulations (which are inherently backreaction free). An important step towards answering the question is therefore to study the mean redshift drift versus the drift of the mean redshift in a model universe based on a Newtonian N-body simulation. The results of such a study will also be presented here.
\newline\newline
Section \ref{sec:dz} below introduces the theoretical framework for computing the redshift drift in a Newtonian N-body simulation and discusses the importance of the optical drift. Numerical results from computing the redshift drift along light rays propagated through the N-body model are presented in section \ref{sec:results} while section \ref{sec:conclusion} provides a summary and concluding remarks.

\section{Redshift drift in an inhomogeneous universe}\label{sec:dz}
Several formalisms for computing the redshift drift in inhomogeneous cosmological models have been proposed within the previous few years \cite{dz_BianchiI_2, optical_effects, bigOnLight, nonlinearities, Asta_first_dz, Asta_cosmography}. The redshift drift will here be computed using the formalism presented in \cite{Asta_cosmography}, i.e. using the decomposition
\begin{align}
&\delta z =\\ \nonumber & \delta \tau_0 E_e \int_{\lambda_e}^{\lambda_0} d \lambda \left(  -\kappa^{\mu} \kappa_{\mu} +\Sigma^{\bf O}    +  e^\mu \Sigma^{\bf e}_\mu    +       e^\mu   e^\nu \Sigma^{\bf ee}_{\mu \nu} + e^{\mu}\kappa^{\nu}\Sigma^{\bf e\kappa}_{\mu\nu} \right) ,
\end{align}
with
\begin{align}\label{eq:components}
&\Sigma^{\bf O} :=  - \frac{1}{3} u^\mu u^\nu R_{\mu \nu}     + \frac{1}{3}D_{\mu} a^{\mu} + \frac{1}{3} a^\mu a_\mu    \\   
&\Sigma^{\bf e}_\mu  :=  - \frac{1}{3}  \theta a_\mu    -   a^{ \nu} \sigma_{\mu \nu}  + 3 a^{ \nu} \omega_{\mu \nu}    - h^{\nu}_{\mu} \dot{a}_\nu     \\   
&\Sigma^{\bf ee}_{\mu \nu} :=     a_{ \mu}a_{\nu }  + D_{  \mu} a_{\nu }    -  u^\rho u^\sigma  C_{\rho \mu \sigma \nu}   -  \frac{1}{2} h^{\alpha}_{\,\mu} h^{\beta}_{\, \nu}  R_{ \alpha \beta }   \\
& \Sigma^{\bf e\kappa}{\mu\nu} :=2(\sigma_{\mu\nu}-\omega_{\mu\nu})\\
&\kappa^{\mu} = h^{\mu}_{\nu}\dot e^{\nu}.   
\end{align}
Quantities are defined in the usual manner with $R_{\mu\nu}$ the Ricci tensor, $C_{\rho\mu\sigma\nu}$ the Weyl tensor, $D_{\mu}$ the 3D spatial covariant derivative, $h_{\mu\nu}$ projecting onto spatial hypersurfaces orthogonal to the velocity field $u^{\mu}$ with a corresponding acceleration $a^{\mu}$. The velocity field is associated with an expansion scalar $\theta$, vorticity $\omega_{\mu\nu}$ and shear $\sigma_{\mu\nu}$. The 4-vector $e^{\mu}$ is the spatial direction vector of the light ray as seen by the observer with velocity $u^{\mu}$. Triangular brackets indicate the trace-free symmetric part of the spatial projection. The convention $c=1$ is used throughout.

\begin{figure}
\centering
\includegraphics[scale = 0.6]{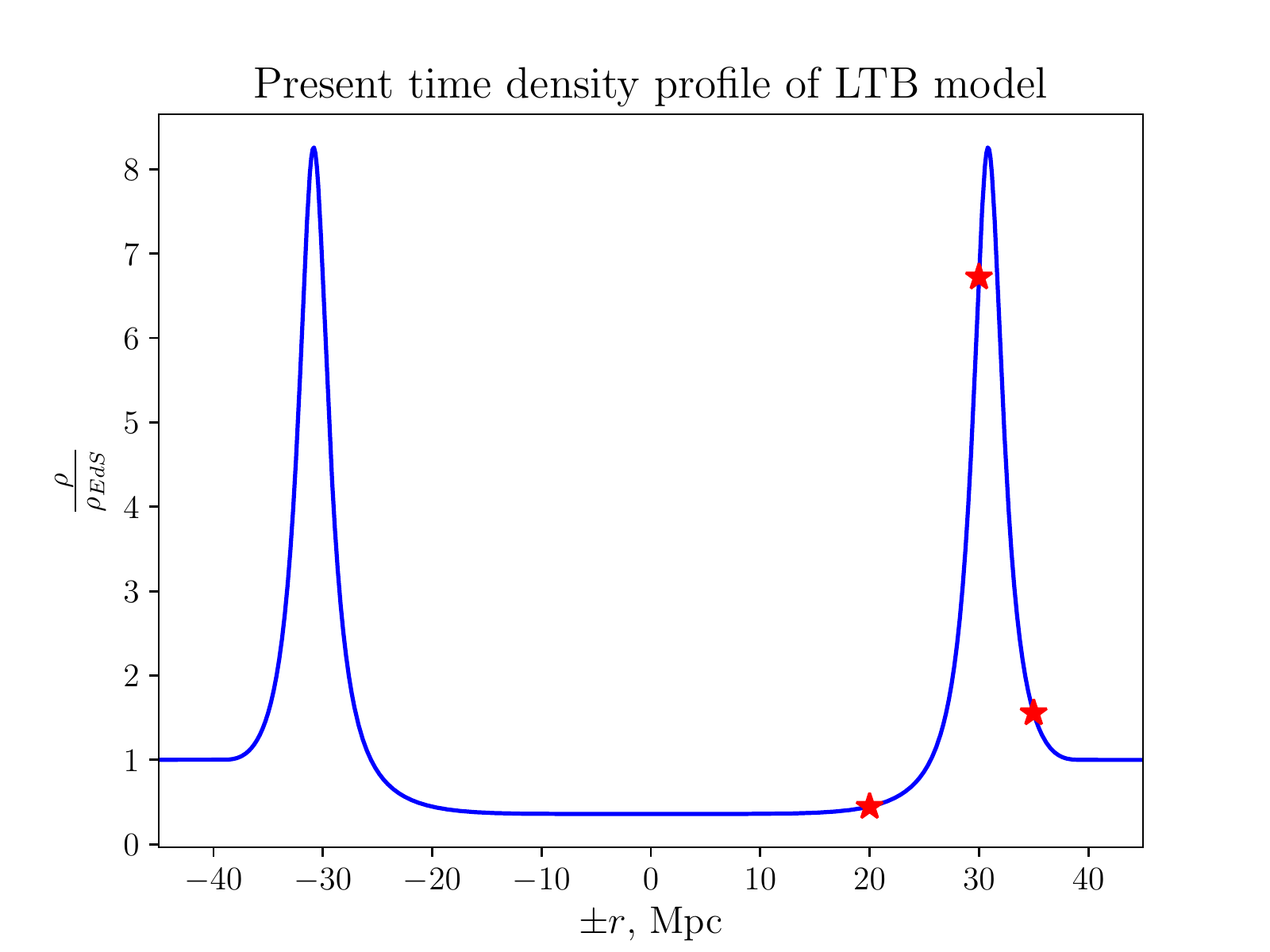}
\caption{Density profile of considered LTB model with stars marking observer positions.}
\label{fig:density_stars}
\end{figure}
\begin{figure*}
	\centering
	\subfigure[]{
		\includegraphics[scale = 0.5]{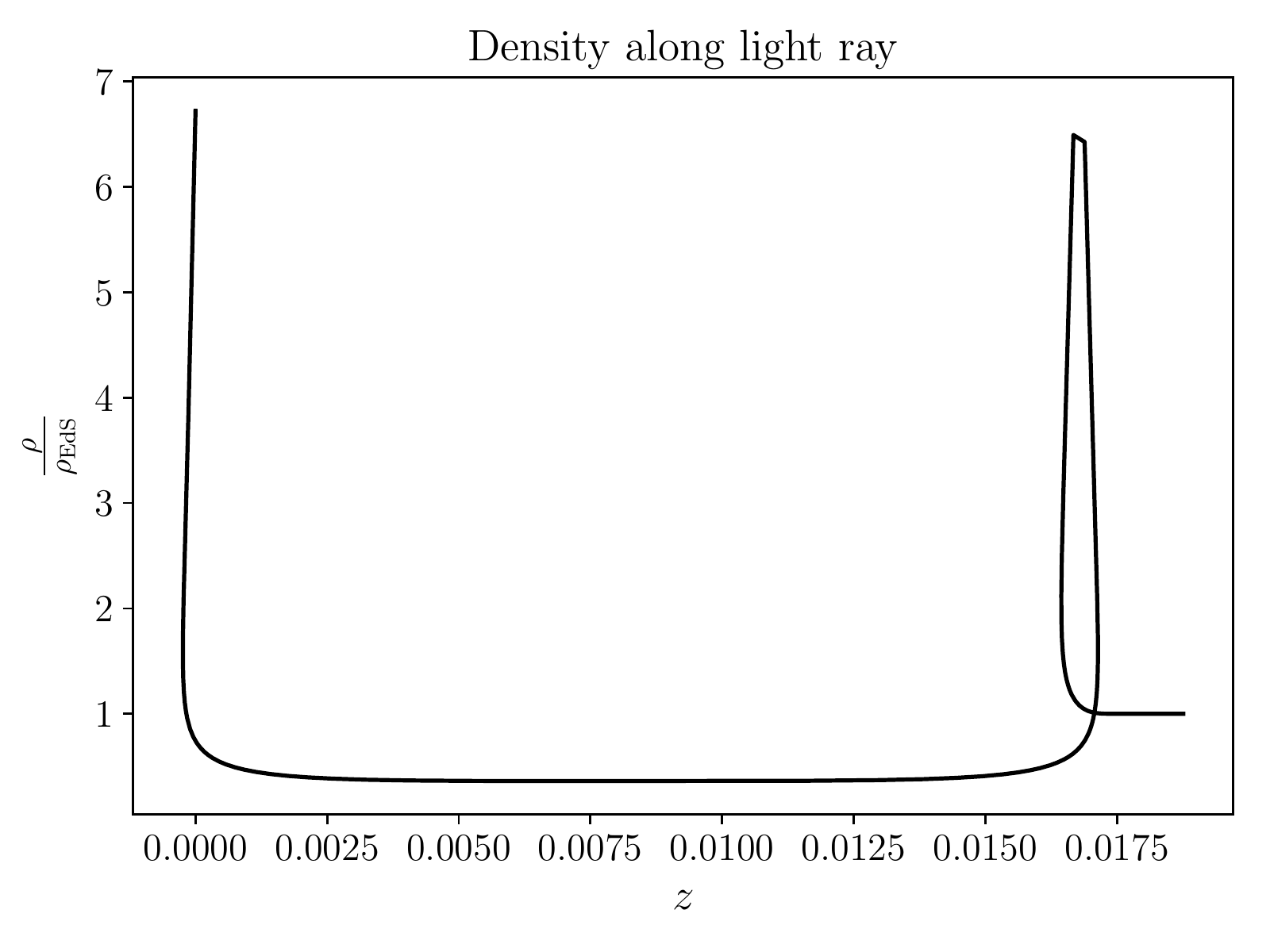}\label{subfig:rho1}
	}
	\subfigure[]{
		\includegraphics[scale = 0.5]{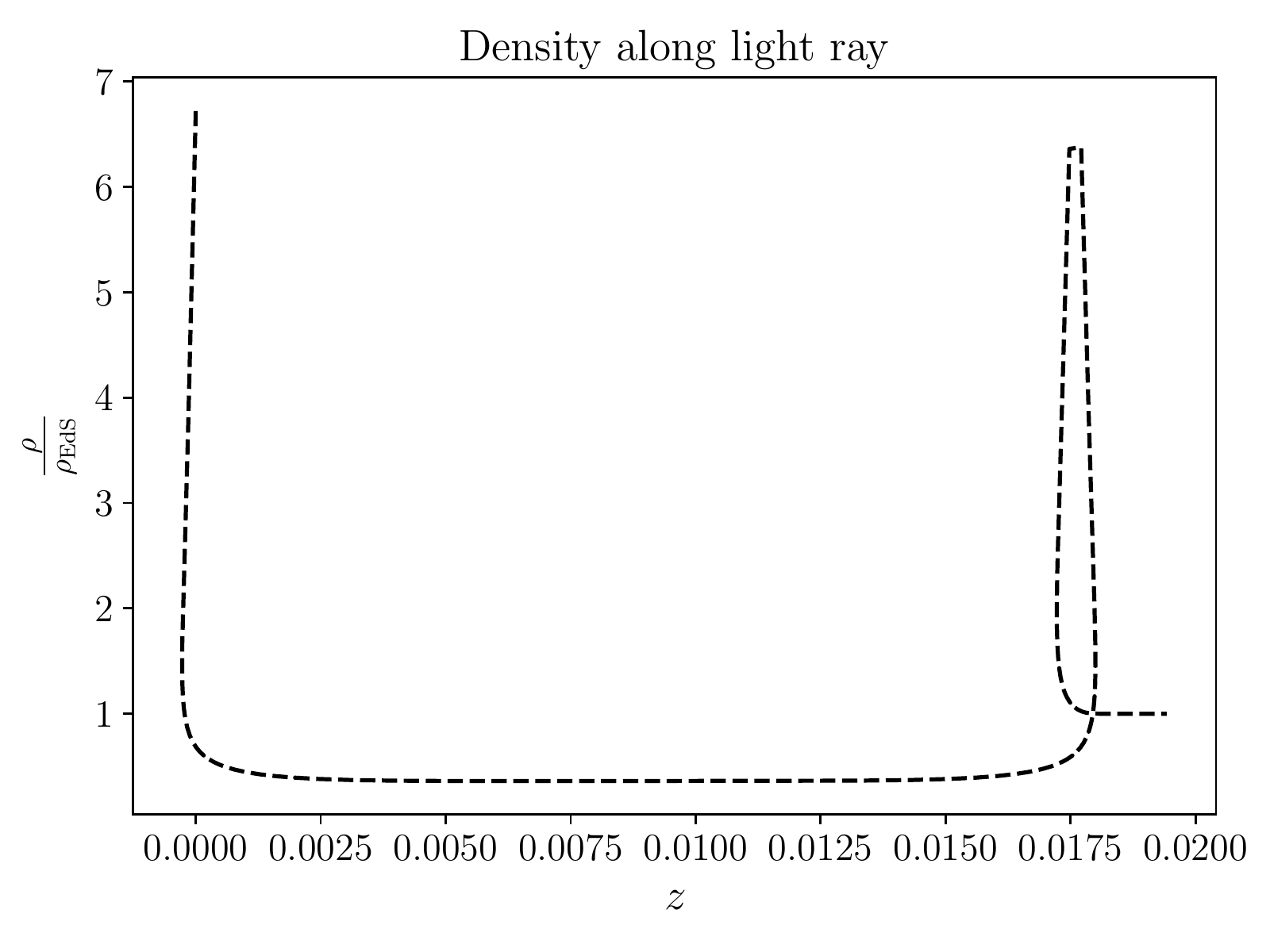}\label{subfig:rho2}
	}\par
	\subfigure[]{
		\includegraphics[scale = 0.5]{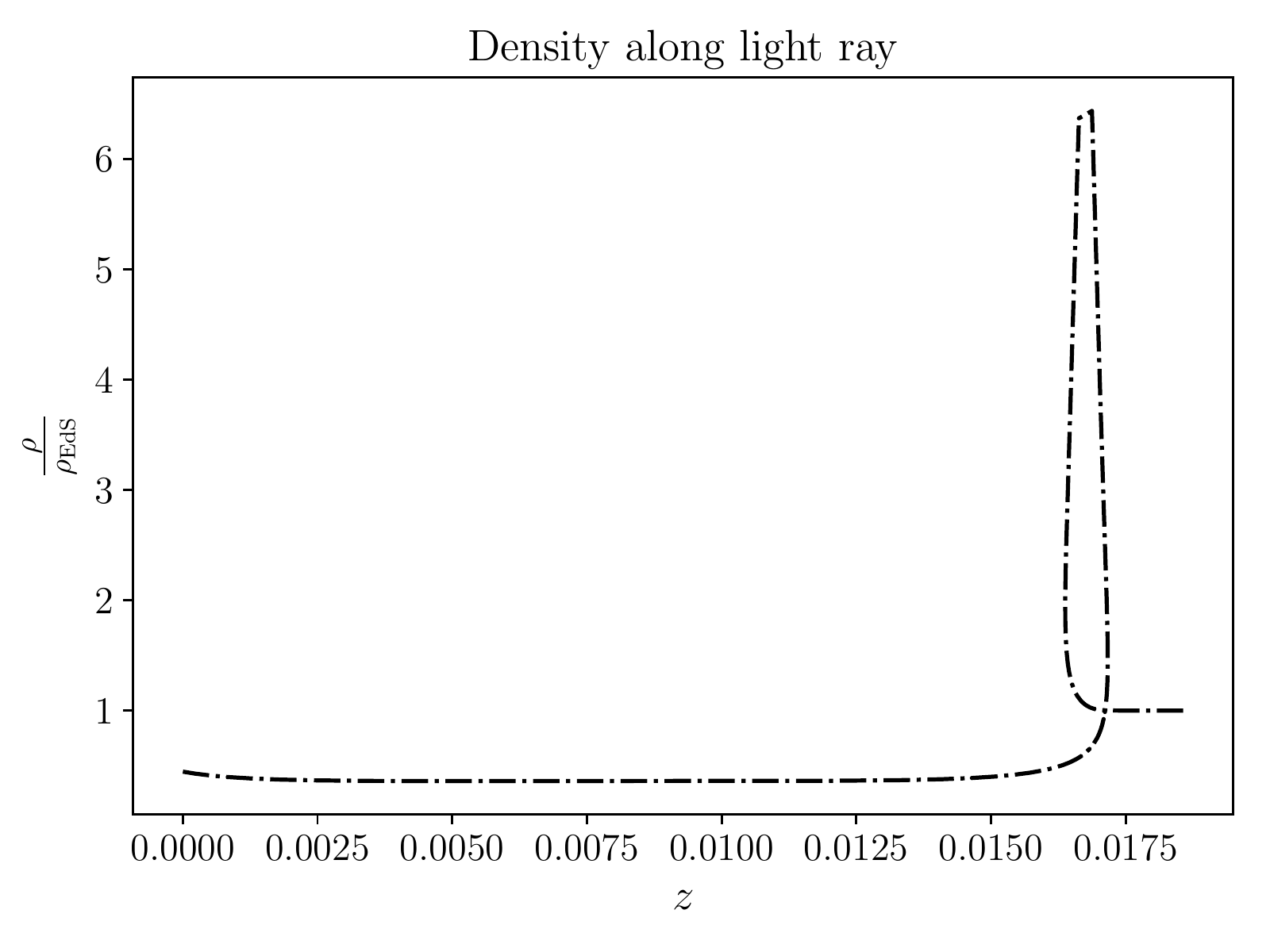}\label{subfig:rho3}
	}
	\subfigure[]{
		\includegraphics[scale = 0.5]{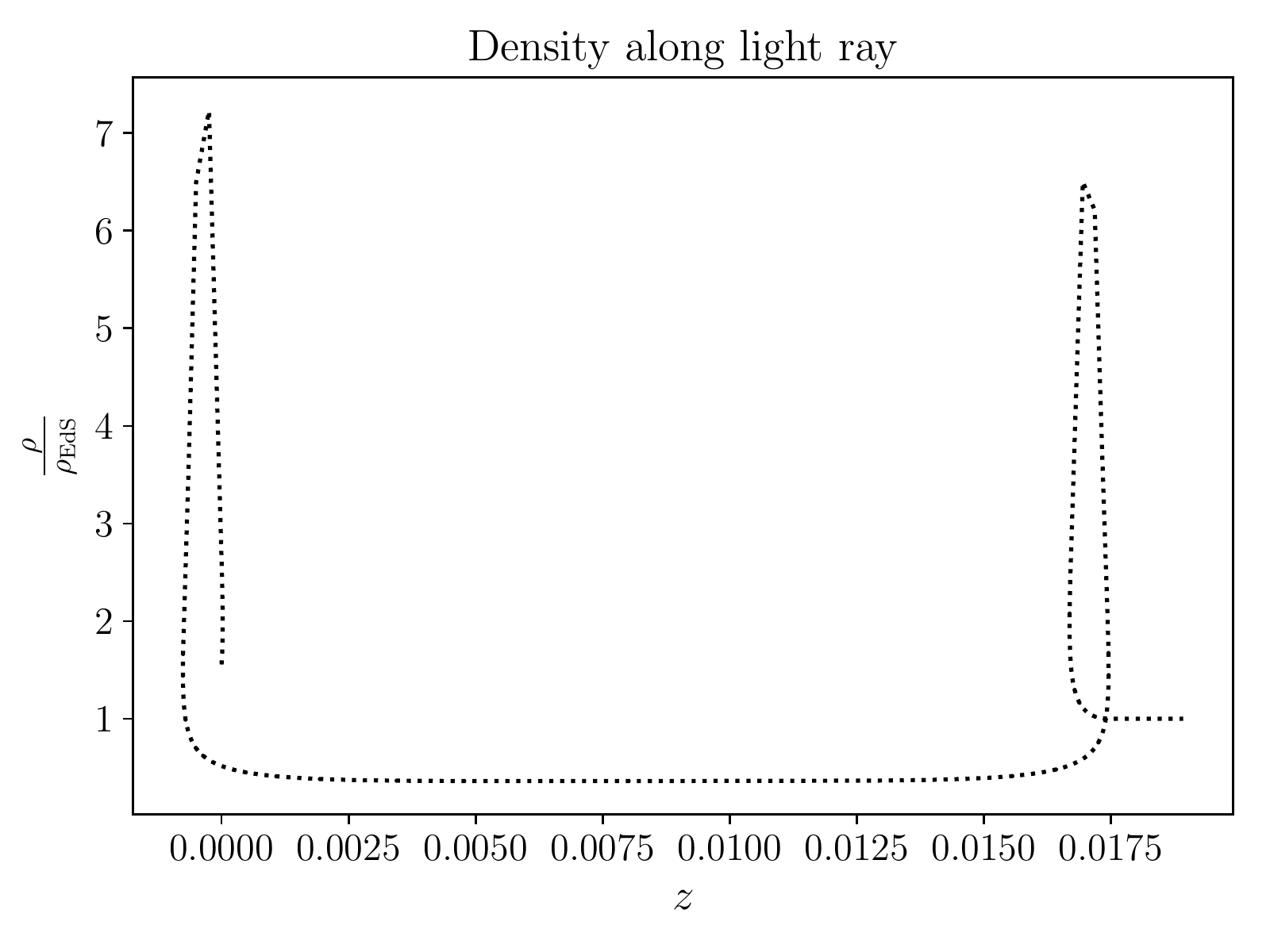}\label{subfig:3ho4}
	}
	\caption{Density along the four considered light rays. Line type is chosen to fit with the line type for $\delta z$ in figure \ref{fig:components} as well as the projections shown in figure \ref{fig:P}
		.}
	\label{fig:density}
\end{figure*}

\subsection{Comment on optical drift}\label{subsec:optical}
\begin{figure*}
	\centering
	\subfigure[]{
		\includegraphics[scale = 0.5]{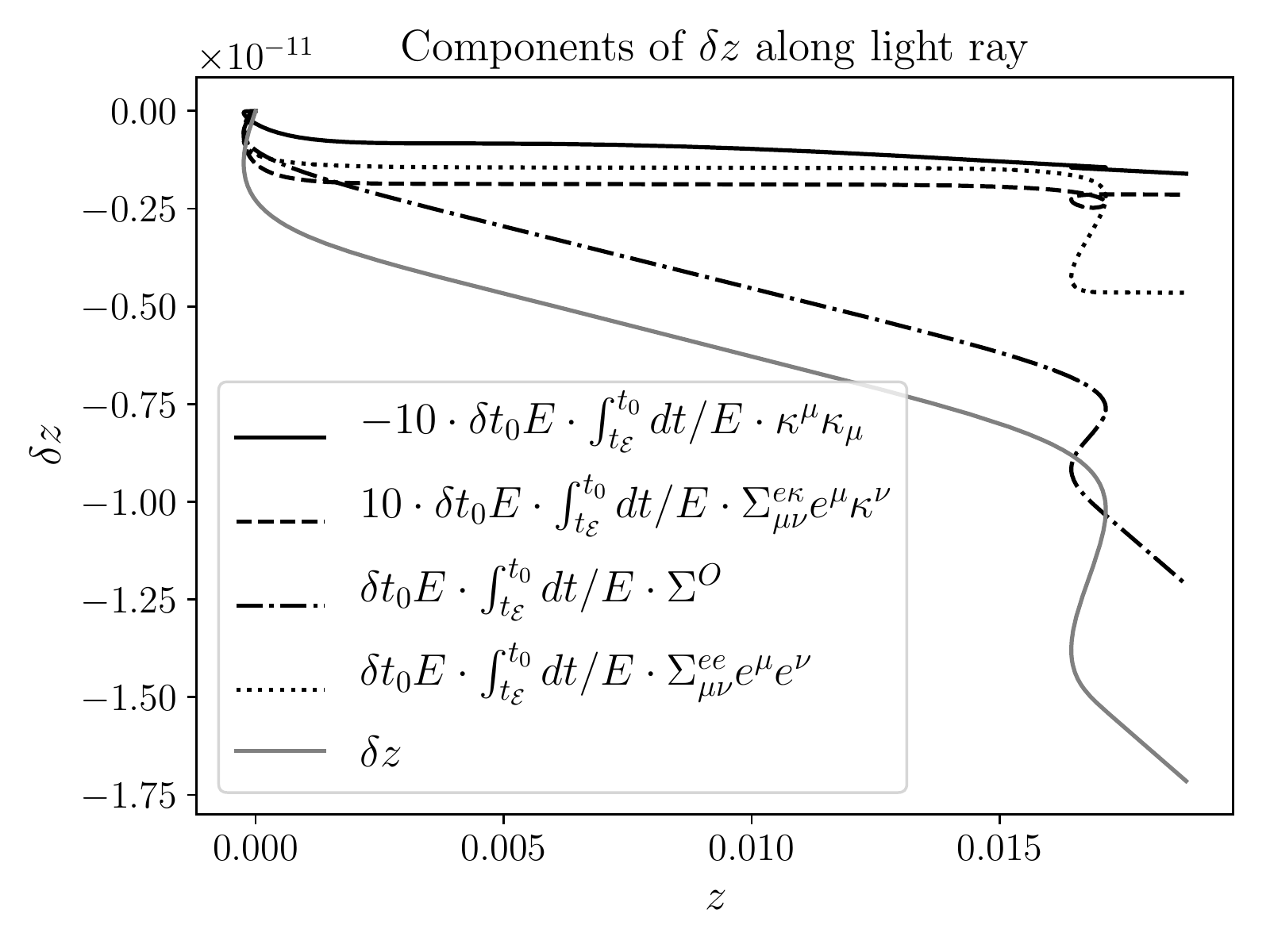}\label{subfig:comp1}
	}
	\subfigure[]{
		\includegraphics[scale = 0.5]{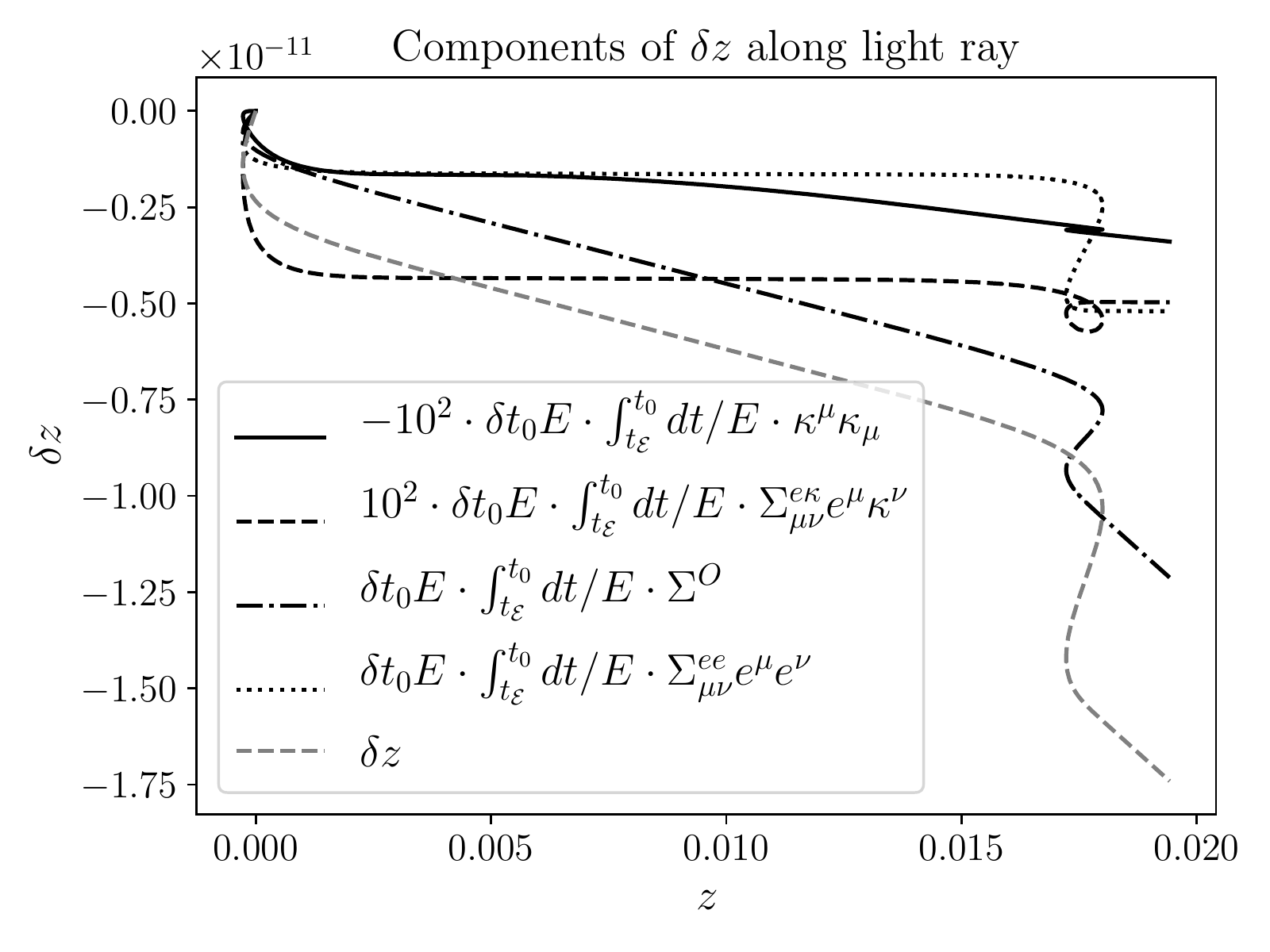}\label{subfig:comp2}
	}\par
	\subfigure[]{
		\includegraphics[scale = 0.5]{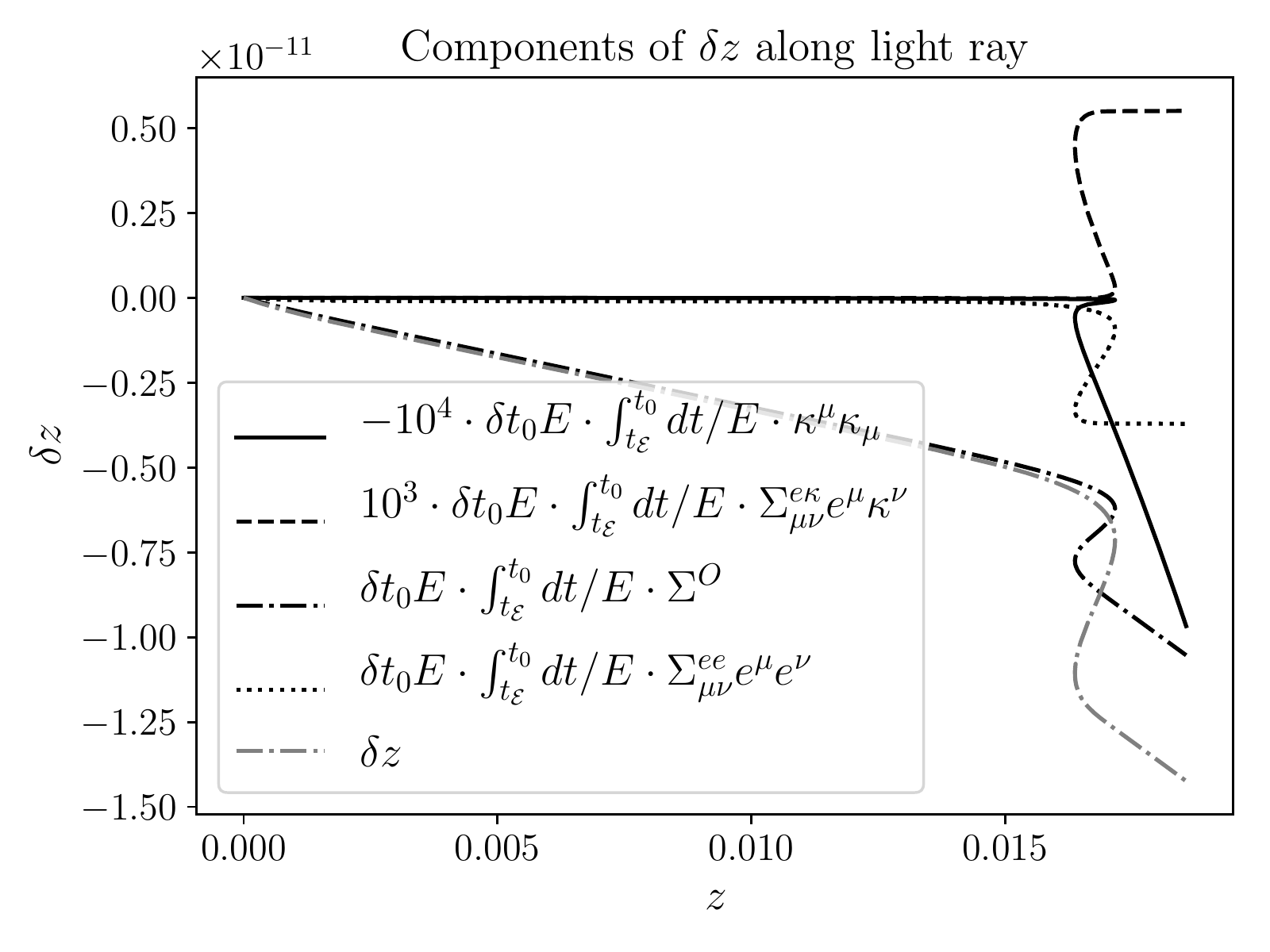}\label{subfig:comp3}
	}
	\subfigure[]{
		\includegraphics[scale = 0.5]{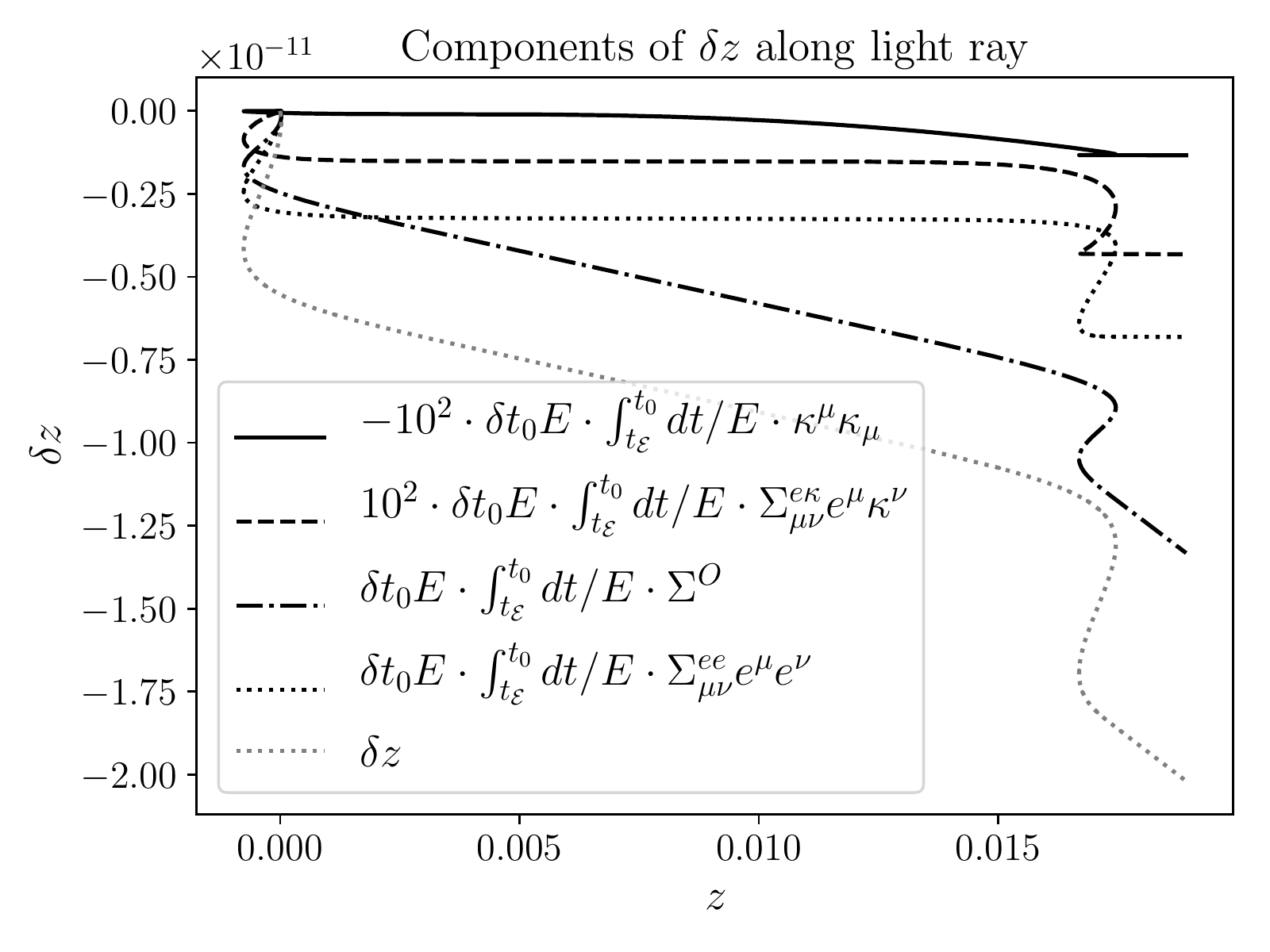}\label{subfig:comp4}
	}
	\caption{Redshift drift and its components along the four considered light rays. The line type of $\delta z$ was chosen to fit with the line type used for the density plots in figure \ref{fig:density} as well as the projections shown in figure \ref{fig:P} such that a given line type of $\delta z$, density and projection correspond to the same light ray.}
	\label{fig:components}
\end{figure*}

As discussed in \cite{dz_LTB3}, computations of the redshift drift simplify tremendously if the optical drift (quantified by $\kappa^{\mu}$) can be neglected (see e.g. \cite{dz_LTB3} for a discussion on the optical drift and \cite{parallax, position_drift1, position_drift2, position_drift3} for related considerations). The optical drift will therefore be neglected for the computations of the redshift drift along light rays through the N-body simulation studied here. To justify this, note that it was in \cite{dz_LTB3} found that for Lemaitre-Tolman-Bondi (LTB) models \cite{LTB1,LTB2,LTB3} with a central void surrounded by an overdensity and with the observer placed in the FLRW region outside the double structure, the optical drift could be neglected to a relative precision of $10^{-3}$ when computing the redshift drift. However, the observers in a Newtonian N-body simulation will in general be placed in an inhomogeneous patch of spacetime and not in an FLRW background. Thus, to further judge the significance of neglecting the optical drift, the redshift drift in the LTB model used in \cite{dz_LTB3} is here studied with observers placed at different positions inside the void to verify that the optical drift still represents a subdominant contribution to the redshift drift. 
\newline\indent
Four non-radial light rays have been studied with 3 different observers with non-radial lines of sight. Figure \ref{fig:density_stars} shows the present-day density profile of the LTB model with stars marking the positions of the observers. Two different lines of sight have been considered for the observer represented by the middle star.
\newline\indent
The density fields along the light rays are shown in figure \ref{fig:density}, and in figure \ref{fig:components}, the different contributions to the redshift drift are shown. Note that $\Sigma^{\bf e\kappa}_{\mu\nu}$ is identically zero in the LTB spacetime and therefore not included in the plots. The figures show that even for an observer in the inhomogeneous LTB region, the terms involving the optical drift are clearly subdominant. However, for one particular observer and line of sight, the optical drift becomes as large as roughly of order 10 \% based on an overall estimate along the light ray. This indicates that the optical drift cannot be expected to be entirely negligible for the simulation data. This must be kept in mind when considering the results presented further below regarding the redshift drift in the simulated model universe. It must also be kept in mind that the study presented in this section is not a thorough systematic study, but merely an initial study tentatively suggesting that the optical drift represents a sub-dominant contribution to the redshift drift along most lines of sight even if the observer is placed inside a structure.
\newline\indent
\noindent
Details regarding the LTB computations and a remark on the stabilization of the effect of the optical drift around 10\% can be found in appendix \ref{app:P} together with a discussion of emitter velocity.

\subsection{Formalism for computing the redshift drift in a Newtonian N-body simulation}
To apply the above formalism to simulation data from GADGET-2 \cite{Gadget,Gadget2}, snapshots from the simulation must be combined with a spacetime metric. According to \cite{GreenWaldMap} this can be done by introducing the perturbed line element
\begin{align}
	ds^2 = -(1+2\psi)dt^2 + a^2(1-2\psi)\left( dx^2+dy^2+dz^2\right) ,
\end{align} 
where $\nabla^2 \psi = 4\pi Ga^2\left(\rho-\rho_{\rm bg} \right) $ with $\rho_{\rm bg}$ the background matter density of the simulation and $\rho$ the actual local matter density. The densities are obtained by TSC (triangular shaped cloud) interpolation of the particle masses in each snapshot. The metric function $\psi$ is then found using FFTW3\footnote{http://www.fftw.org}.
\newline\indent
Using this line element, the geodesic equation can be solved for light rays propagating through the simulation box. Quadri-linear interpolation is used to interpolate within and between snapshots.
\newline\newline
With the line element as given above, we can compute the integrand components in equation \ref{eq:components}. For simplicity, the optical drift is neglected. In addition, the shear, vorticity and acceleration all vanish at background order so to first order in these quantities the integrand components reduce to 
\begin{align}
&\Sigma^{O} \equiv  - \frac{1}{3} u^\mu u^\nu R_{\mu \nu}    \\   
&\Sigma^{e}_\mu  \equiv   - \frac{1}{3}  \theta a_\mu      \\   
&\Sigma^{ee}_{\mu \nu} \equiv    -  u^\rho u^\sigma  C_{\rho \mu \sigma \nu}    . 	
\end{align}
In the last line it was also utilized that $ \frac{1}{2} h^{\alpha}_{\,\mu} h^{\beta}_{\, \nu}  R_{ \alpha \beta } $ vanishes identically.
\newline\indent
The total expression for the redshift drift is thus given by the integral over just 3 terms: The projections of the Ricci tensor, $R_{\mu\nu}$, and the Weyl tensor, $C_{\rho\mu\sigma\nu}$, along the light path as well as the projection of the acceleration scaled by the expansion rate, i.e.
\begin{widetext}
\begin{align}
\delta z \approx E_e\delta \tau_0\int_{t_e}^{t_0}\frac{dt(1+2\psi)}{E^{(n)}}\left[-\frac{1}{3}u^{\mu}u^{\nu}R_{\mu\nu}- \frac{1}{3}  \theta a_\mu e^\mu-u^{\rho}u^{\sigma}e^{\mu}e^{\nu}C_{\rho\mu\sigma\nu} \right].
\end{align}
\end{widetext}
Note that the energy $E^{(n)}$ under the integral is computed using the velocity field normal to the spatial hypersurfaces i.e. $n^{\mu} = -(1+2\psi)\partial_{\mu}t = (-(1+2\psi),0,0,0)$ and not the fluid velocity field\footnote{This was pointed out by Asta Heinesen.}. In practice, the difference between these two is of course very small and at first order, it only affects the background part of the Ricci term of the integrand (see e.g. \cite{syksy_light} for details regarding the choice of velocity field and energy computations when integrating over the affine parameter versus over the time coordinate).
\newline\newline
This integral is computed as a finite sum on-the-fly while solving the geodesic equation for the given spacetime, observer and emitter\footnote{The issue regarding emitter velocity discussed in \cite{dz_LTB3} is neglected here. See appendix \ref{app:P} for a justification and discussion.}.
\newline\newline
With the line element given above, the three considered terms contributing to the redshift drift are at lowest order in $\psi$ and its derivatives given by
\begin{align}
-\frac{1}{3}u^{\mu}u^{\nu}R_{\mu\nu} = -\frac{4\pi G}{3}\rho,
\end{align}
\begin{align}
	-\theta a_{\mu}e^{\mu} = 3H\left( \partial_t \delta u^i + H\delta u^{j} \delta_j^i \right)e_{i},
\end{align}
and
\begin{align}\label{eq:weyl_contribution}
	\begin{split}
&	-u^{\rho}u^{\sigma}e^{\mu}e^{\nu}C_{\rho\mu\sigma\nu}  =\\
	& -e^i e^j\left( \partial_i \partial_j \psi + \Gamma^{\gamma(\rm bg)}_{ij}\psi_{,\gamma} -\frac{4\pi G a^2}{3}\left( \rho-\rho_{\rm bg} \right)\delta_{ij}  \right) ,
	\end{split}
\end{align}
where the superscripted ``(bg)'' on the Christoffel symbol indicates that only the background values are used. The 4-vector $e^{\mu}:=u^{\mu} - k^{\mu}/E$ is the spatial projection of the null geodesic tangent vector as seen by an observer comoving with the fluid and is to be understood as given at background order everywhere except for in the last term. $\delta u^{\mu}$ represents the perturbation to the velocity field of the fluid which is obtained by TSC interpolation of the velocity field, $v^i$, from the snapshot. The full 4-velocity field is given by $u^\mu\propto(1,v^i)$, with normalization according to $u^{\mu}u_{\mu} = -1$.
	
\subsection{Simulation setup}
The simulation data was obtained by running GADGET-2 \cite{Gadget,Gadget2} with initial conditions generated with N-GenIC\footnote{https://wwwmpa.mpa-garching.mpg.de/gadget}. The Einstein-de Sitter (EdS) model with a reduced Hubble parameter of 0.7 was chosen as the simulation background. This choice will give slightly enhanced quantitative results compared to a $\Lambda$CDM model where only 30\% of the energy content is inhomogeneous.
\newline\indent
The simulation was run with $512^3$ particles in a box with side lengths $512{\rm Mph}/h$ and 24 snapshots in the interval roughly corresponding to $z\in [0,1]$ were produced.
\begin{figure*}
	\centering
	\subfigure[]{
		\includegraphics[scale = 0.5]{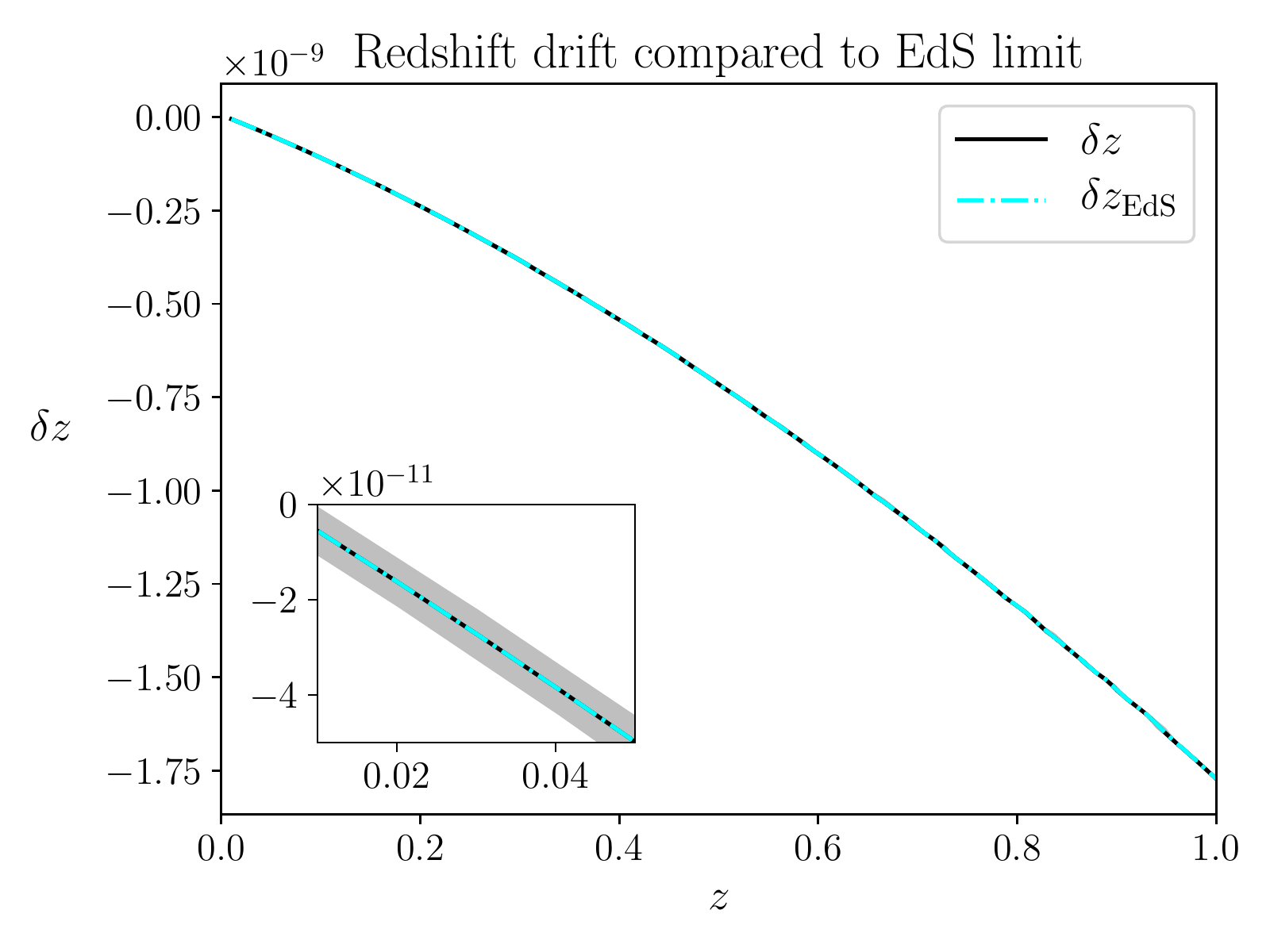}\label{subfig:dz}
	}
	\subfigure[]{
		\includegraphics[scale = 0.5]{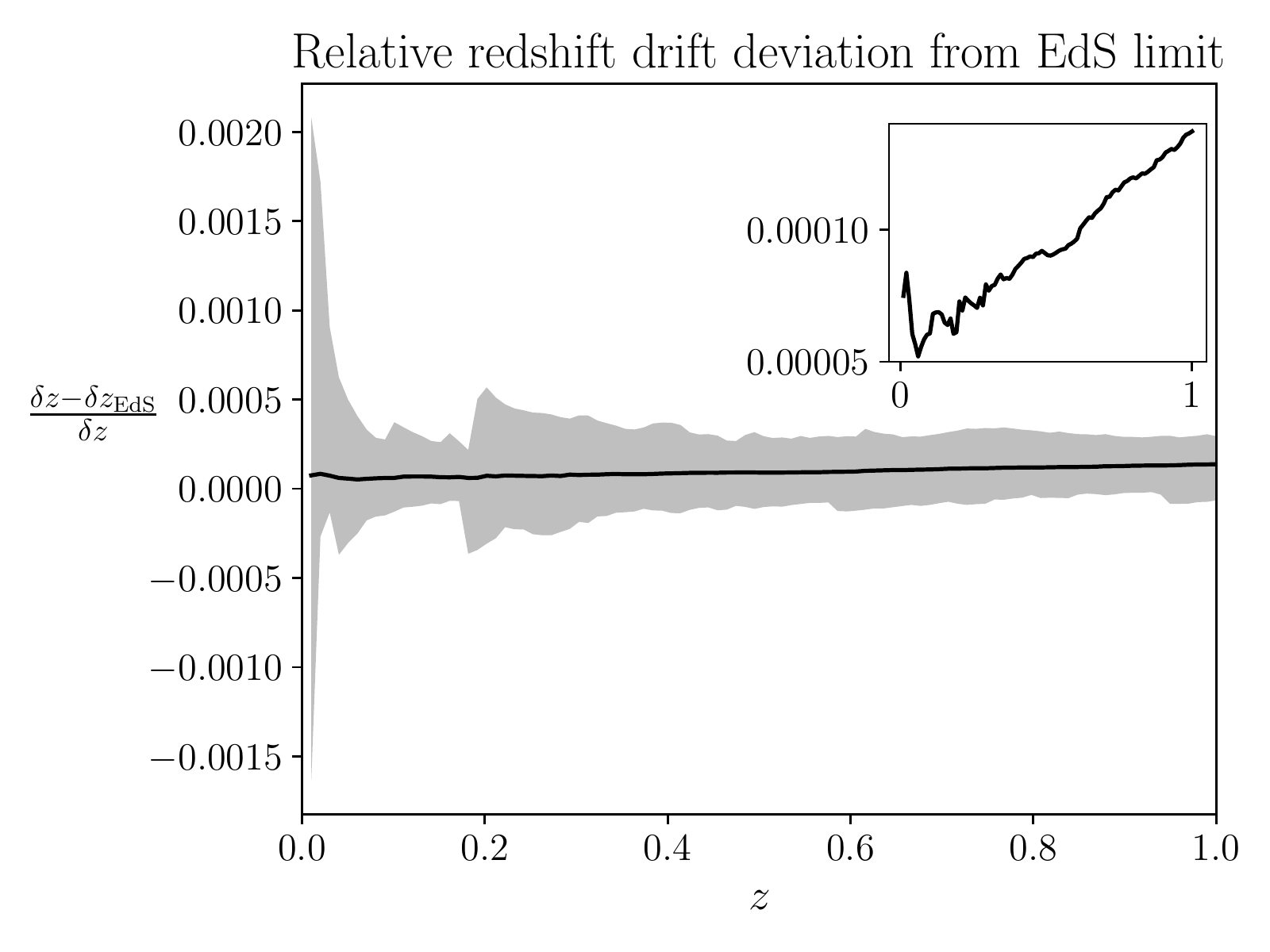}\label{subfig:ddz}
	}
	\caption{Mean and fluctuations in the redshift drift along 189 light rays. In subfigure \ref{subfig:dz} the redshift drift is shown together with the redshift drift of the EdS model but the two lines are indistinguishable. The close-up in the subfigure is included to show the fluctuations around the mean since these are too small to be visible in the ordinary plot. Subfigure \ref{subfig:ddz} shows the relative deviation between the redshift drift along the light rays compared to the EdS redshift drift. A close-up is included to show the deviation of the mean from being exactly zero.}
	\label{fig:189_dz}
\end{figure*}	
\section{Numerical results}\label{sec:results}
\begin{figure}
	\centering
	\subfigure[]{
		\includegraphics[scale = 0.5]{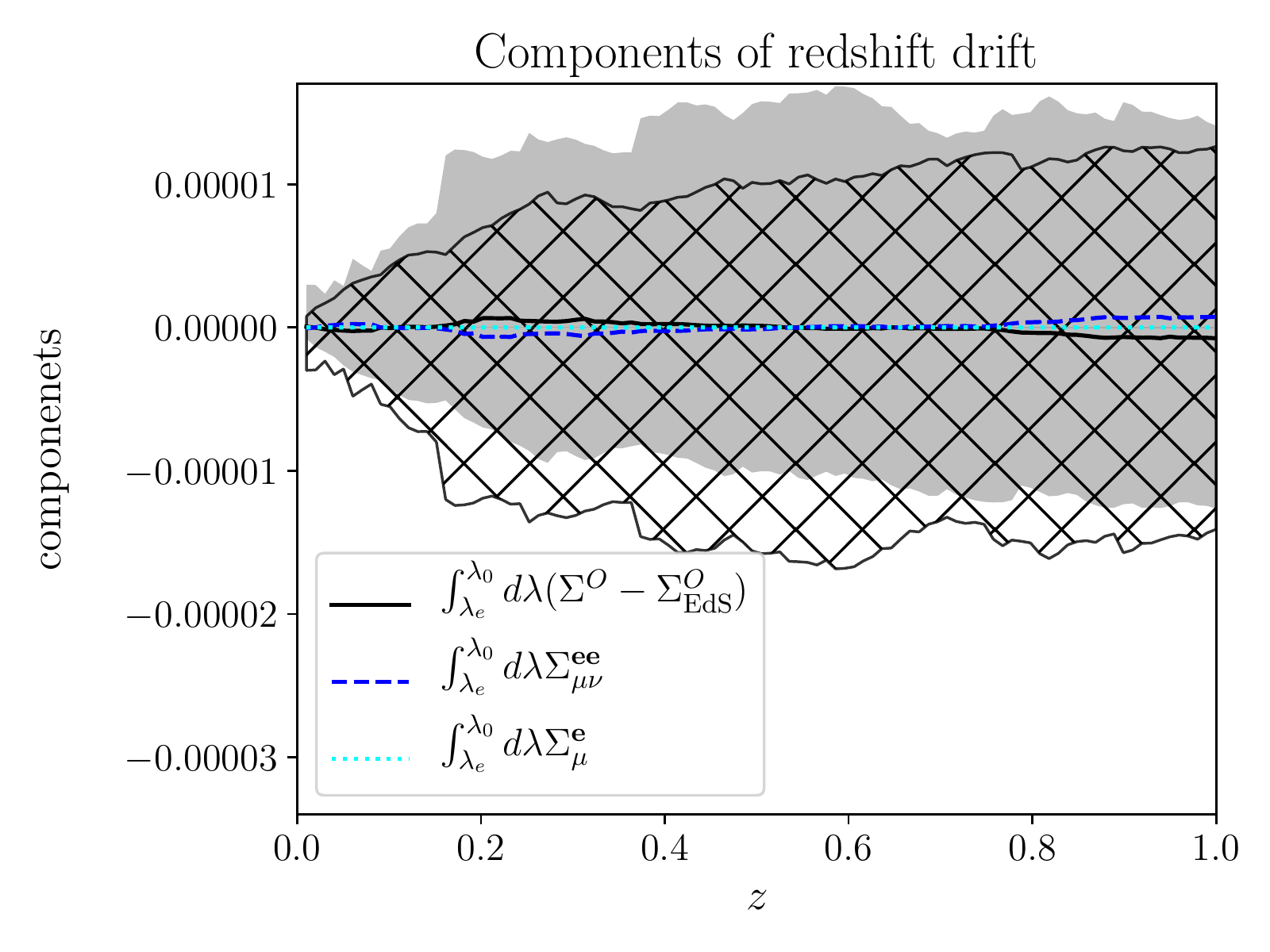}
	}
	\caption{Mean and fluctuations of the individual components of the redshift drift along 189 light rays. The gray shaded area represents the fluctuations of the $\Sigma^{O}$ component while the hatched area indicates the fluctuations of the $\Sigma^{\bf ee}_{\mu\nu}$ component. The fluctuations of the acceleration term are too small to be visible in the figure.}
	\label{fig:189_components}
\end{figure}
\begin{figure}
	\centering
	\subfigure[]{
		\includegraphics[scale = 0.5]{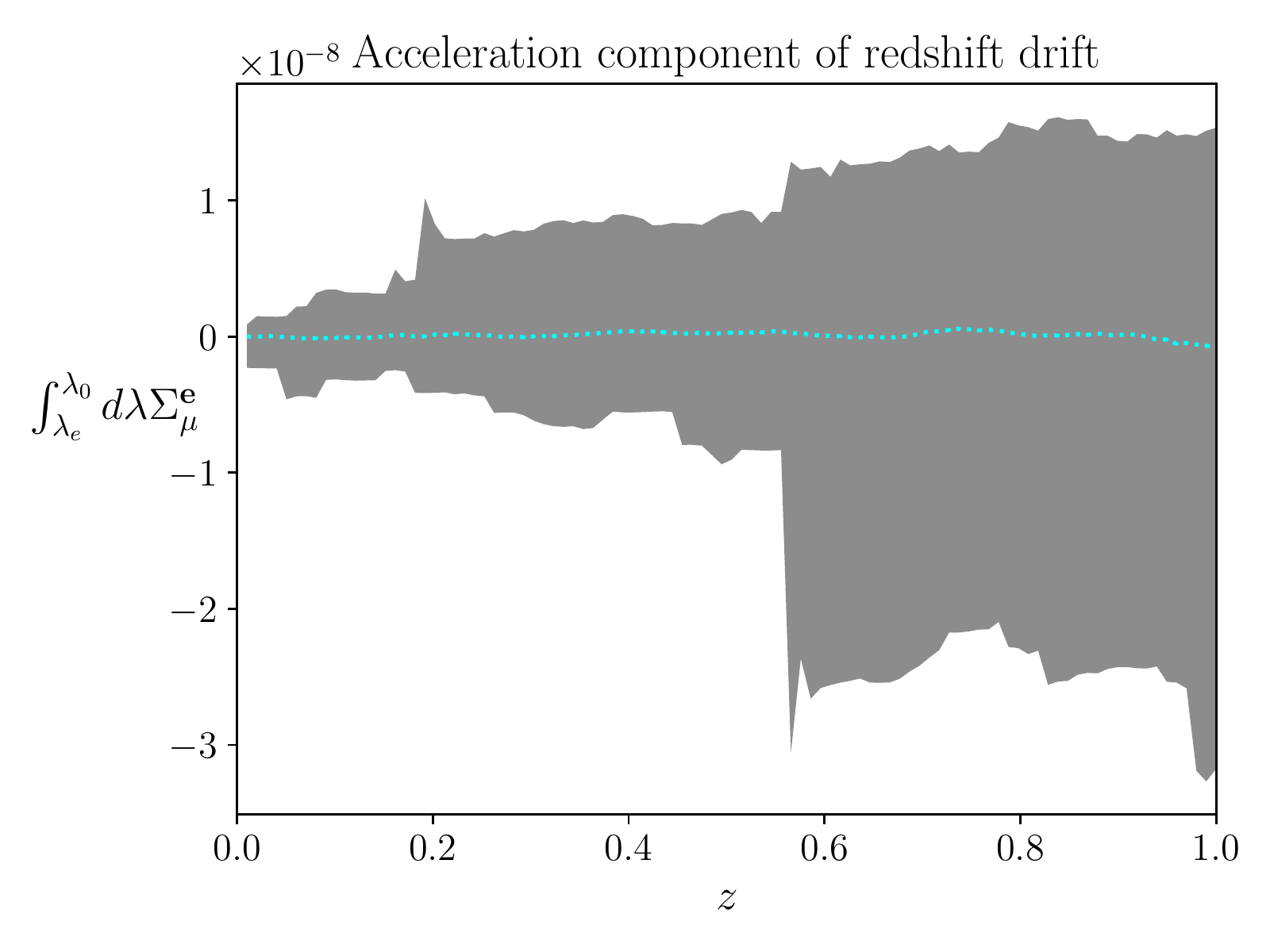}
	}
	\caption{Mean and fluctuations of the acceleration component, $\int_{\lambda_e}^{\lambda_0}d\lambda\Sigma^{\rm e}_{\mu}$, of the redshift drift along 189 light rays.}
	\label{fig:189_acc}
\end{figure}
189 light rays were considered, all with random lines of sight and observers placed at random spatial positions (all at present time). Observation time, $\delta t_0$, was set to 30 years. The main results are shown in figure \ref{fig:189_dz} which shows the redshift drift along the light rays compared to the redshift drift in the EdS model. The relative deviation between the mean redshift drift and the redshift drift of the EdS model (which represents the drift of the mean redshift) is of order $10^{-4}$. Considering that the redshift drift itself is of order $10^{-9}$, this means that the deviation is of absolute order $10^{-13}$ which is roughly around the expected precision of the computations.
\newline\indent
Figure \ref{fig:189_components} shows the mean and fluctuations of the individual components of the redshift drift. The acceleration term is too small to be clearly visible and is therefore shown independently in figure \ref{fig:189_acc} where it is seen that the fluctuations in the acceleration term is of order $10^{-8}$. The mean of this term is of order $10^{-10}$ which is several orders of magnitude below the other two terms. In relation to this, it should be noted that earlier studies \cite{Linder_dz1,Linder_dz2, acc_dz, variance_dz} show that the full non-linear peculiar acceleration contribution to the redshift drift signal can be larger than the cosmic signal along individual light rays but that the signal should average out when considering many rays in different directions. The small contribution from the acceleration found here is presumably due to the peculiar acceleration being suppressed since it was computed from a smoothed velocity field representing large scale structures. Thus, the current study cannot meaningfully contribute to the quantification of effects from peculiar acceleration.
\newline\indent
\begin{figure*}
	\centering
	\subfigure[]{
		\includegraphics[scale = 0.5]{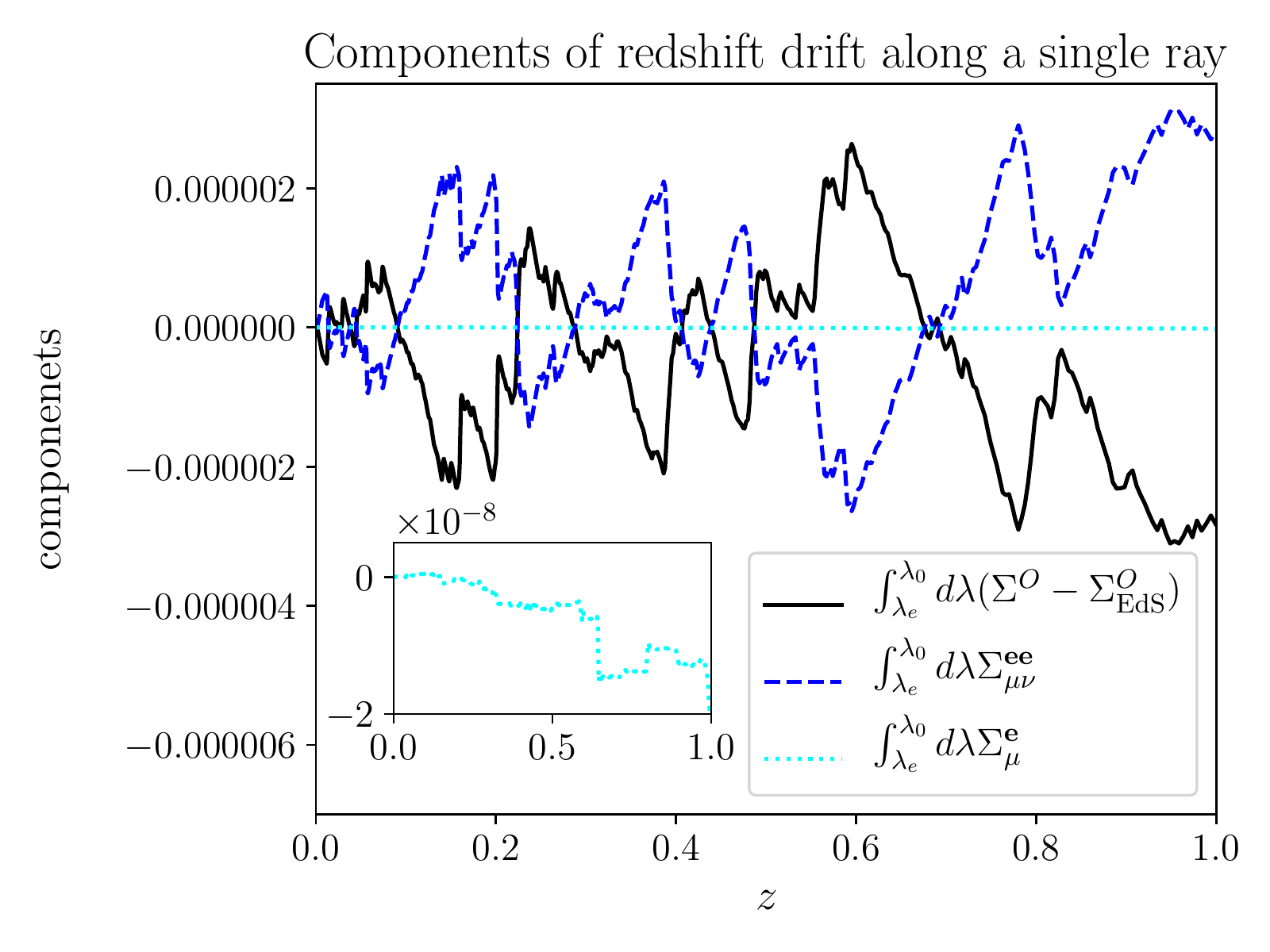}\label{subfig:comp_single}
	}
	\subfigure[]{
		\includegraphics[scale = 0.5]{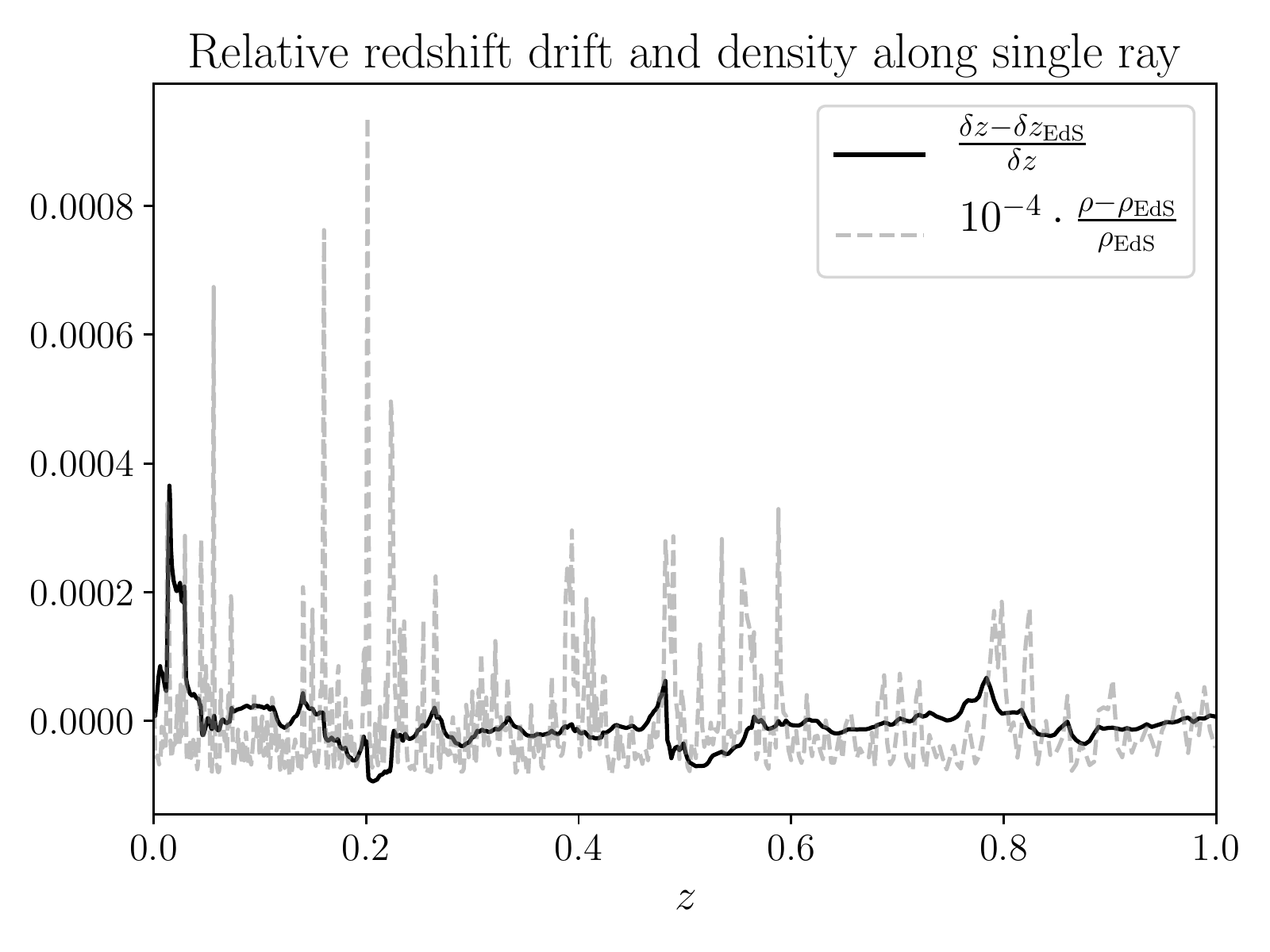}\label{subfig:rel_single}
	}
	\caption{Redshift drift along a random light ray. Subfigure \ref{subfig:comp_single} shows the individual components of the redshift drift while subfigure \ref{subfig:rel_single} shows the relative deviation between the redshift drift along the light rays and the EdS light ray. The relative redshift drift is plotted on top of the density distribution along the light ray.}
	\label{fig:single}
\end{figure*}
Figure \ref{fig:189_components} shows that the mean of the contribution from the two dominant (i.e. non-acceleration) perturbative terms cancel each other almost exactly in the mean computation, leaving only the contribution representing the background redshift drift. A similar symmetry is seen in the contours of the shaded and hatched areas indicating the fluctuations of the two components. Figure \ref{fig:single} shows the redshift drift along a random light ray and supports this symmetry: It is clearly seen that the two dominant perturbative contributions cancel each other to high precision at all points along the light ray, indicating that the dominant term in equation \ref{eq:weyl_contribution} is the term proportional to the density fluctuation. This is reminiscent of the cancellations between the shear and expansion rate fluctuations in the redshift discussed in \cite{dr}. Indeed, in \cite{dr} one can see that the redshift along light rays in the studied N-body simulation is always very close to the background redshift with a deviation of at most of order $10^{-5}$. It is therefore not surprising that the redshift drift is also close to the background value everywhere along the light rays. However, in agreement with earlier studies \cite{tardis,cheese,syksy_pert}, it was in \cite{dr} found that the redshift along individual light rays in Swiss-cheese models based on LTB structures was also everywhere close to the background redshift, with deviations of order $10^{-3}$. But the results of \cite{dz_LTB2} are {\em not} that the redshift drift is always close to the background redshift drift in LTB models.
\newline\indent
To establish the relationship between these cancellation relations along light rays and their possible relation to cancellations of e.g. the terms of the kinematical backreaction \cite{kinematical,kinematical2}, it could be interesting to consider models of averaging along light cones such as the setups presented in \cite{lightcone, null_cone}. In addition, one intriguing point with the results discussed above is the significance of the Weyl tensor and the question of under what conditions its redshift drift contribution (nearly) cancels with the contribution from the fluctuation in the Ricci tensor -- more specifically, if there is a connection between the electric part of the Weyl tensor being dominated by the Ricci curvature, and other aspects of a spacetime. It may be interesting to study this more thoroughly, e.g. with inspiration from the analysis of the ``quiet'' spacetime presented in \cite{quiet}.

\section{Summary and conclusions} \label{sec:conclusion} 
The redshift drift was computed along 189 light rays through a simulated spacetime obtained with the Newtonian N-body simulation code GADGET-2 run with an Einstein-de Sitter background. It was found that the two main non-background contributions to the redshift drift cancel almost exactly along the light rays and therefore the mean redshift drift equals the drift of the mean redshift to high precision. This is contrary to earlier results based on inhomogeneous cosmological models. Specifically, it was in \cite{dz_BianchiI_2, another_look} found for specific models that exhibit significant cosmic backreaction that the mean redshift drift in these models deviates significantly from the drift of the mean redshift. These combined results support the work presented in \cite{dz_test3} where an observational signal of cosmic backreaction was devised based on the notion that the mean redshift drift deviates from the drift of the mean redshift in a spatially statistically homogeneous and isotropic universe only if there is significant cosmic backreaction.
\newline\indent
The cancellation further means that the local fluctuations in the redshift drift along individual light rays are very modest. The results presented here thus indicate that future redshift drift measurements only will contain a small bias/error due to fluctuations from structures. While this is good news, such a conclusion is too strong based on the limits of the considered N-body simulation; the density and velocity fields of the simulation are smoothed to represent large scale structures and therefore the importance of e.g. peculiar acceleration of the emitter cannot be realistically determined from the simulation. Indeed, while the contribution to the redshift drift from peculiar acceleration was here found to be several orders of magnitude smaller than the other contributions, earlier work \cite{Linder_dz1,Linder_dz2, acc_dz, variance_dz} indicates that the peculiar acceleration may (for some types of sources) be even larger than the cosmological contribution to the redshift drift. Presumably, this does not spoil the ambition to measure the cosmic redshift drift since the contribution from the peculiar acceleration according to these studies should become negligible upon averaging over several points of observation.
\newline\indent
It is lastly noted that the results presented here are based on inhomogeneous cosmological models with EdS backgrounds and thus the small fluctuations found here should be expected to correspond to even smaller fluctuations in the redshift drift in similar models with a $\Lambda$CDM background.

\begin{acknowledgments}
The author is funded by the Carlsberg Foundation. The author thanks Asta Heinesen for pointing out an error in the original expression for the redshift drift. Some of the computations done for this project were performed on the UCloud interactive HPC system managed by the eScience Center at the University of Southern Denmark.
		
\end{acknowledgments}

\appendix

\section{Redshift drift computations in an LTB model}\label{app:P}
\begin{figure*}
	\centering
	\subfigure[]{
		\includegraphics[scale = 0.5]{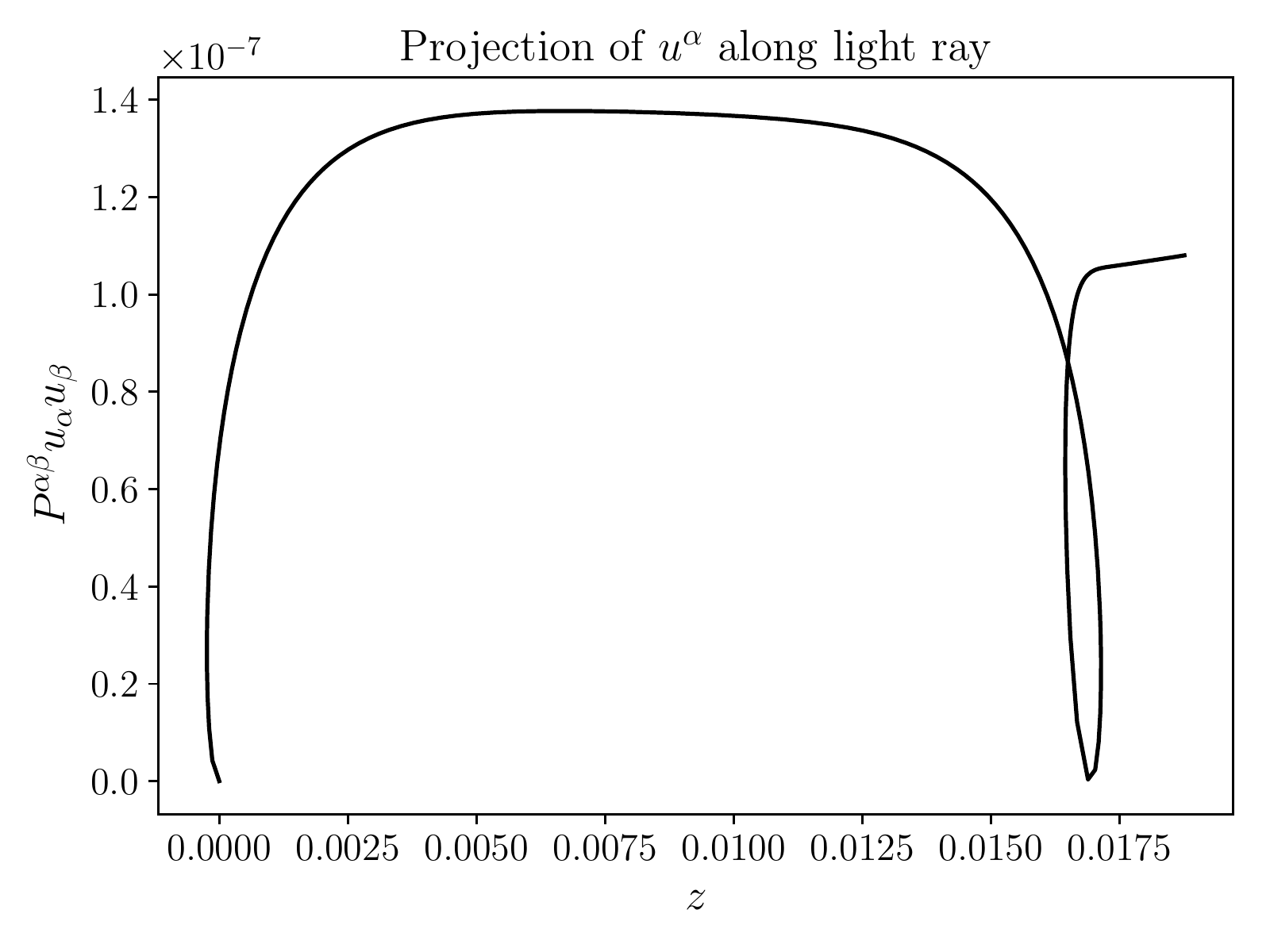}\label{subfig:P1}
	}
	\subfigure[]{
		\includegraphics[scale = 0.5]{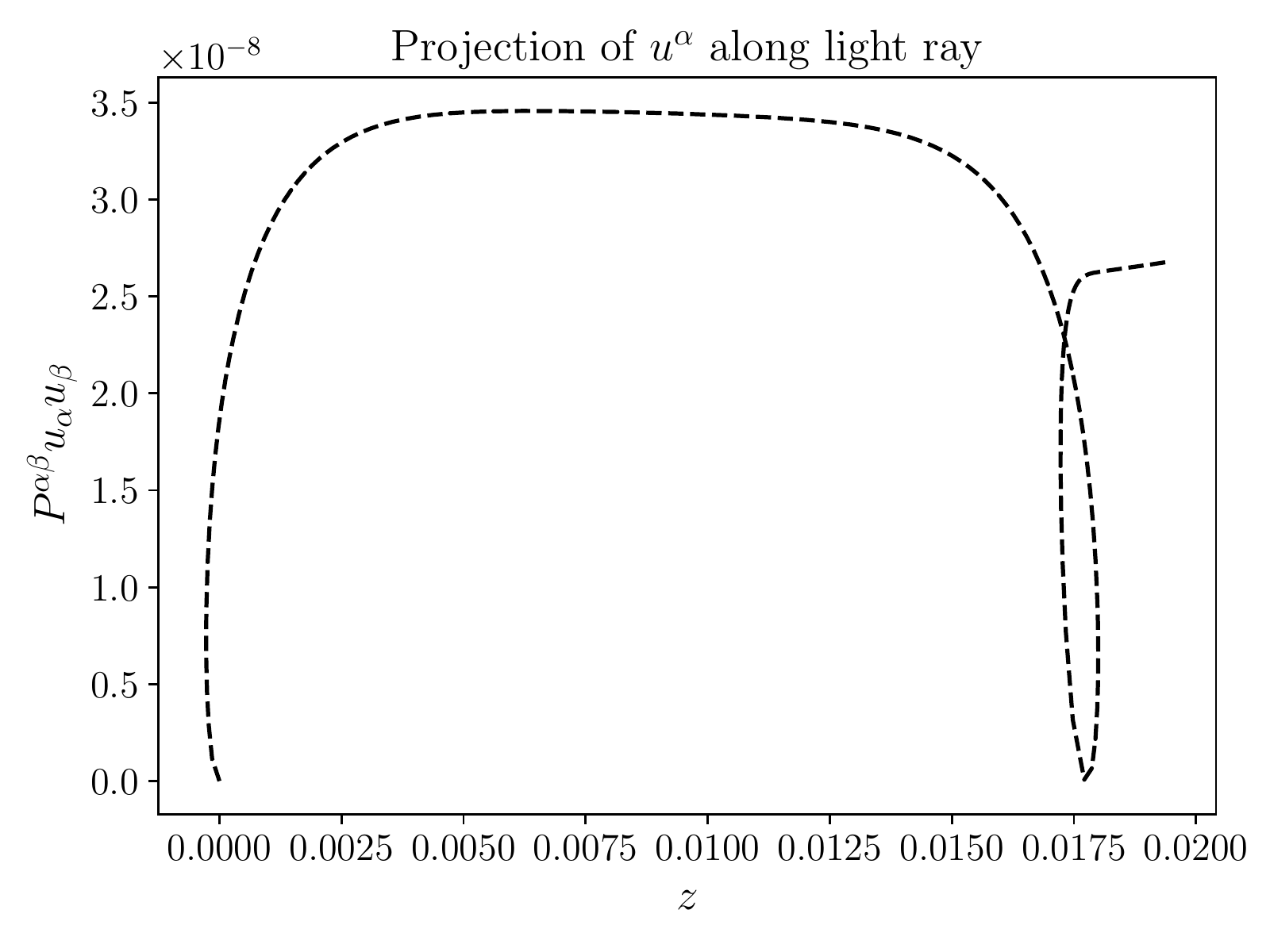}\label{subfig:P2}
	}\par
	\subfigure[]{
		\includegraphics[scale = 0.5]{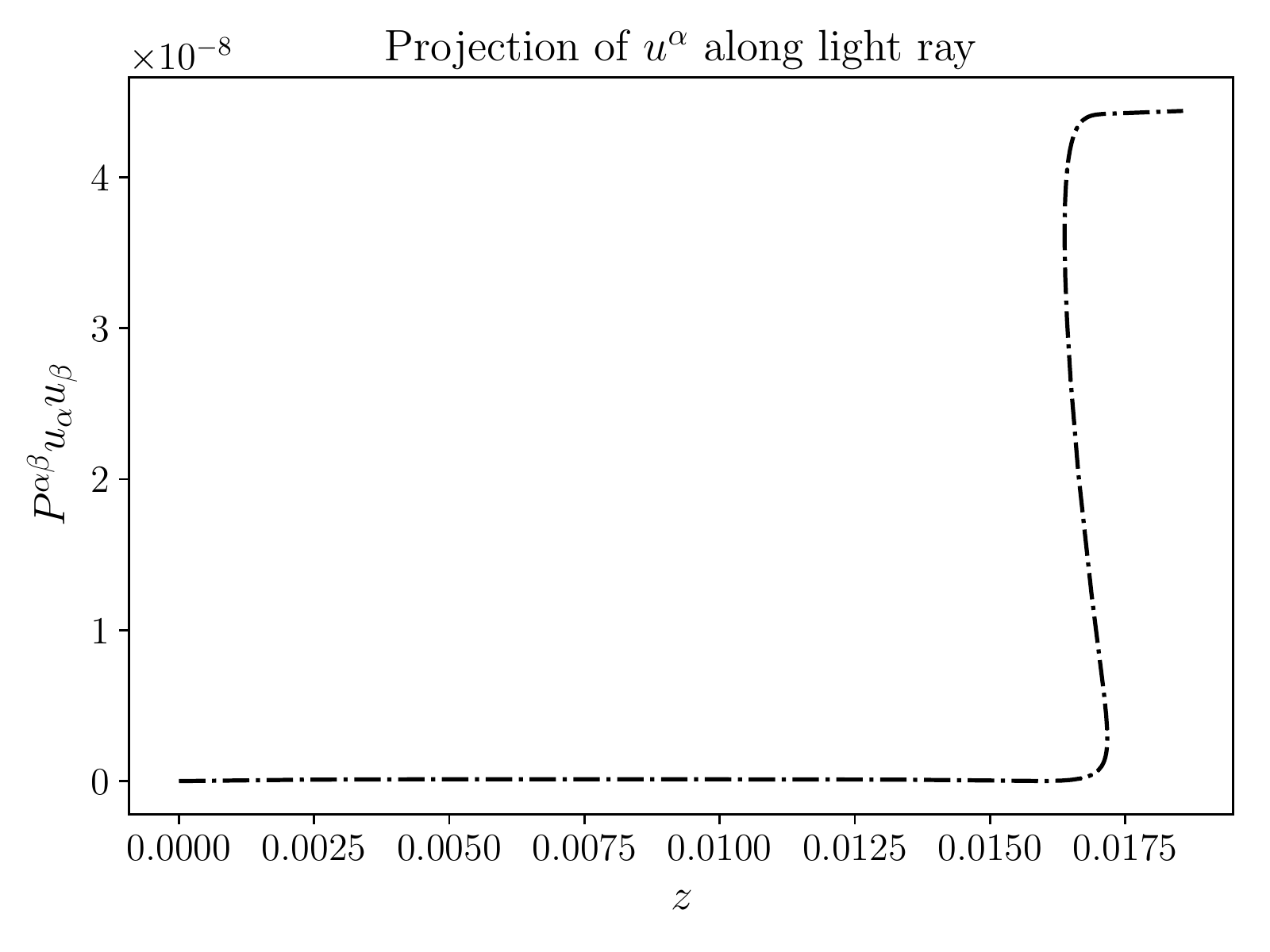}\label{subfig:P3}
	}
	\subfigure[]{
		\includegraphics[scale = 0.5]{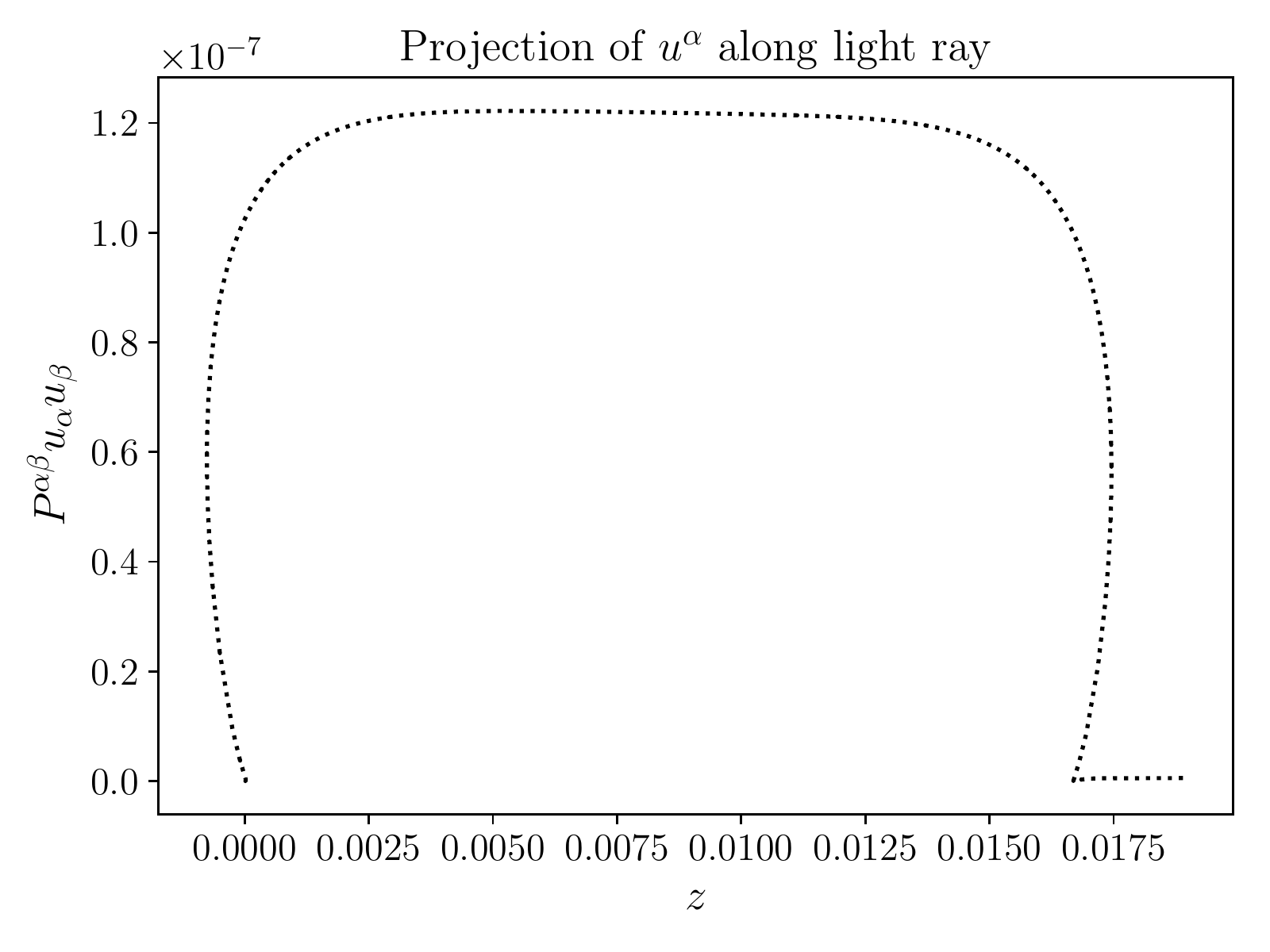}\label{subfig:P4}
	}
	\caption{Projection of comoving velocity field onto the screen space orthogonal to $k^{\mu}$ and the velocity field of a possible emitter (i.e. $P^{\mu}_{\nu} = \delta^{\mu}_{\nu}-\frac{k^\mu k_\nu}{E_m^2}+\frac{k^{\mu}m_{\nu}}{E_m} + \frac{m^{\mu}k_{\nu}}{E_m}$, where $m^\mu$ is a possible velocity field of the emitter and $E_m = -m^\mu k_\mu$). Line types are chosen to fit with the line types used for $\delta z$ in figure \ref{fig:components} and for the densities in figure \ref{fig:density} so that the density, projection and $\delta z$ with the same line types represent quantities along the same individual light ray.}
	\label{fig:P}
\end{figure*}
This appendix serves to give details regarding the LTB model studied in section \ref{subsec:optical}.
\\\\
The line element of the LTB model can be written as
\begin{align}
	ds^2 = -dt^2 + \frac{A_{,r}^2(t,r)}{1-k(r)}dr^2+A^2(t,r)d\Omega^2.
\end{align}
To specify the model used in \cite{dz_LTB3} and the main text here, it is required that the LTB model has a homogeneous big bang which is most naturally ensured by setting $A(t=0, r) = 0$. The model is then further specified by setting 
\begin{align}
	    k(r) = 
	\begin{cases}
	-1.3 \times 10^{-7} r^2\left( \left( \frac{r}{r_b} \right)^6 - 1 \right)^6 ,& \text{if } r < r_b\\
	0 \; ,              & \text{else} \, .
	\end{cases}
\end{align}
As in \cite{dz_LTB3} $r_b$ is chosen to be 40Mpc. Outside the region defined by $r\leq r_b$ the model reduces to the background FLRW model which is here chosen to be the EdS model with reduced Hubble parameter of 0.7. To ensure this, $A(t=t_i, r) = a_ir$ is fixed for an initial time $t_i$ representing the time at which the scale factor of the EdS model is equal to $a_i = 1/1100$. This sets up the initial conditions used for solving the dynamical equation for $A$ which is given by
\begin{align}
	A_{,t}^2 = \frac{2M}{A}-k,
\end{align}
where $M=M(r)$ is fixed by the conditions described above through the considerations presented in \cite{acoleyen}.
\newline\newline
For studying light rays in the LTB model it is necessary to simultaneously solve the geodesic equations and propagation equations for the partial derivatives of the tangent vector, $k^\mu$,
\begin{align}
	\frac{d}{d\lambda}\left(g_{\alpha\beta}k^{\beta} \right) &= \frac{1}{2}g_{\mu\nu,\alpha}k^{\mu}k^{\nu}\\
	\frac{d}{d\lambda}k^{\mu}_{,\nu} &= \frac{\partial}{\partial x^{\nu}}\frac{dk^{\mu}}{d\lambda} - k^{\beta}_{,\nu}k^{\mu}_{,\beta},
\end{align}
where the latter set of equations were presented in \cite{ishak}.
\newline\newline
Naturally, initial conditions are needed in order to solve the above set of equations along light rays. When setting these, the expressions for the components of $\kappa^{\mu}$ will be used and are therefore shown here. By defining $R:=A_{,r}^2/(1-k)$, the components can be written as
\begin{widetext}
	\begin{align}
	\begin{split}
	{\kappa}^t &= 0\\
	{\kappa}^r &= -\frac{1}{k^t}\left[ \left( 1-R\frac{\left( k^r\right) ^2}{\left( k^t\right) ^2} \right)\left(k^r_{,t} +\frac{R_{,t}}{2R}k^r \right) -\frac{k^rk^{\theta}}{\left( k^t\right) ^2}A^2\left( k^\theta_{,t} + \frac{A_{,t}}{A}k^\theta\right) -\frac{k^rk^\phi}{\left( k^t\right) ^2}A^2\sin^2(\theta)\left(k^{\phi}_{,t} +\frac{A_{,t}}{A}k^\phi \right)     \right] \\
	{\kappa}^\theta & = -\frac{1}{k^t}\left[ -\frac{k^\theta k^r}{(k^t)^2}R\left(k^r_{,t} + \frac{R_{,t}}{2R}k^r \right) + \left(1-\frac{(k^{\theta})^2}{(k^t)^2}A^2 \right)\left(k^\theta_{,t} + \frac{A_{,t}}{A}k^{\theta} \right) -\frac{k^{\theta}k^\phi}{(k^t)^2}A^2\sin^2(\theta)\left(k^\phi_{,t} + \frac{A_{,t}}{A}k^\phi \right)   \right]\\
	{\kappa}^{\phi}& = -\frac{1}{k^t}\left[  -\frac{k^\phi k^r}{(k^t)^2}R\left( k^r_{,t} +\frac{R_{,t}}{2R} \right) -\frac{k^\phi k^\theta}{(k^t)^2}A^2\left( k^\theta_{,t} +\frac{A_{,t}}{A}k^\theta\right) +\left(1-\frac{\left( k^{\phi}\right) ^2}{(k^t)^2}A^2\sin^2(\theta) \right)\left(k^\phi_{,t} + \frac{A_{,t}}{A}k^\phi \right)       \right] .
	\end{split}
	\end{align}
\end{widetext}
Utilizing the spherical symmetry of the LTB models, we can set $k^{\phi} $ and its partial derivatives to zero. This trivially means that $\kappa^\phi = 0$ initially. In addition, initial conditions are also set to fulfill $\kappa^r, \kappa^{\theta} = 0$ initially. This implies that there is no optical drift at the position of the observer which seems a reasonable choice. To further specify initial conditions, the partial derivatives of the null condition as well as the definition $\frac{dk^\nu}{d\lambda}:=k^{\mu}\partial_{\mu}k^\nu$ must also be fulfilled.
\newline\indent
As shown in figure \ref{fig:components}, these initial conditions lead to a subdominant optical drift contribution to the redshift drift. However, to judge the significance of this result it must be considered what the corresponding peculiar velocity field of the emitters along the light rays are. As discussed in \cite{dz_LTB3}, the choice of initial conditions for $k^{\mu}_{,\nu}$ fixes the possible peculiar velocities of the emitter. In principle, this means that the redshift drift must be computed with a new set of initial conditions for each point along a light ray in order to ensure that the peculiar velocity field of the emitter is as desired. However, for the light rays considered in \cite{dz_LTB3}, it was found that the actual emitter velocity field corresponding to a single set of initial conditions could be chosen to be very close to the comoving velocity field (which was the desired velocity field). In figure \ref{fig:P} the same is shown for the four light rays considered in the main text. Specifically, figure \ref{fig:P} shows that the comoving velocity field projected orthogonally to the light ray tangent vector and an actually possible velocity field of the emitter is small -- of order $10^{-8}-10^{-7}$.
\\\\
The four light rays were generated using  the initial conditions $k^{\phi} = 0$, $k^t = -1$, $k^{\theta}=0.001,0.005$ and $k^r$ obtained from the null condition. The observers were placed according to $r_b-r = 5,10,20$Mpc. The observer placed at $r = r_b-10$Mpc is the observer with 2 lines of sight. For the other 2 observers, only initial conditions with $k^\theta = 0.005$ were considered. In all cases, the observation time was set to $\delta t_0 = 30$ years. A fifth light ray with initial $r= r_b-10$Mpc and $k^{\theta} = 0.01$ has also been studied but the results are similar (of same order of magnitude) as for the light ray with initial $r = r_b-10$Mpc and $k^{\theta} = 0.005$ so the results for the former light ray are not shown. The fact that the effect of the optical drift is of same order of magnitude for these two light rays is taken to tentatively indicate a stabilization of the optical drift effect at about 10\% in this type of model.


\begin{thebibliography}{}
\bibitem{tensions1} W. Handley, Curvature tension: evidence for a closed universe, Phys. Rev. D 103, L041301 (2021), arXiv:1908.09139
\bibitem{tensions2} E. Di Valentino, A. Melchiorri and J. Silk, Planck evidence for a closed Universe and a possible crisis for cosmology, Nature Astron. 4 196–203 (2020), arXiv:1911.02087
\bibitem{tensions3} E. Di Valentino, A. Melchiorri and J. Silk Cosmic Discordance: Planck and luminosity distance data exclude LCDM, arXiv:2003.04935
\bibitem{tensions4} L. Perivolaropoulos, F. Skara, Challenges for $\Lambda$CDM: An update, arXiv:2105.05208v3 [astro-ph.CO]
\bibitem{tensions5} Weiqiang Yang et al., Revealing the effects of curvature on the cosmological models, arXiv:2210.09865v1 [astro-ph.CO] 
\bibitem{tensions6} Elcio Abdalla et al., Cosmology Intertwined: A Review of the Particle Physics, Astrophysics, and Cosmology Associated with the Cosmological Tensions and Anomalies, J. High En. Astrophys. 2204, 002 (2022), arXiv:2203.06142v3 [astro-ph.CO]
\bibitem{tensions7} Eleonora Di Valentino, Emmanuel Saridakis, Adam Riess, Cosmological tensions in the birthplace of the heliocentric model, arXiv:2211.05248v1 [astro-ph.CO]
\bibitem{tension_added_sunny_1} Sunny Vagnozzi, Eleonora Di Valentino, Stefano Gariazzo, Alessandro Melchiorri, Olga Mena, Joseph Silk, The galaxy power spectrum take on spatial curvature and cosmic concordance, Phys. Dark Univ. 33 (2021) 100851, arXiv:2010.02230v3 [astro-ph.CO]
\bibitem{tension_added_sunny_2} Sunny Vagnozzi, Abraham Loeb, Michele Moresco, Eppur è piatto? The cosmic chronometer take on spatial curvature and cosmic concordance, Astrophys. J. 908 (2021) 84, arXiv:2011.11645v2 [astro-ph.CO]


\bibitem{sandage} Allan Sandage, The Change of Redshift and Apparent Luminosity of Galaxies due to the Deceleration of Selected Expanding Universes, Astrophysical Journal, vol. 136, p.319 (1962)
\bibitem{mcvittie} G. C. McVittie, Appendix to The Change of Redshift and Apparent Luminosity of Galaxies due to the Deceleration of Selected Expanding Universes, Astrophysical Journal, vol. 136, p.334 (1962)


\bibitem{bolejko_flux} Krzysztof Bolejko, Chengyi Wang, Geraint F. Lewis, Direct detection of the cosmic expansion: the redshift drift and the flux drift, arXiv:1907.04495v1 [astro-ph.CO]
\bibitem{SKA} H.-R. Klockner et al., Real time cosmology - a direct measure of the expansion rate of the Universe, contribution to SKA Cosmology Chapter, Advancing Astrophysics with the SKA (AASKA14), Conference, Giardini Naxos (Italy), June 9th-13th 2014, arXiv:1501.03822
\bibitem{feasible1} J. Liske et al., Cosmic dynamics in the era of Extremely Large Telescopes, Mon.Not.Roy.Astron.Soc.386:1192-1218,2008, arXiv:0802.1532v1 [astro-ph]
\bibitem{feasible2} S. Cristiani, et al., The CODEX-ESPRESSO experiment: cosmic dynamics, fundamental physics, planets and much
more..., Nuovo Cim.B122:1159-1164,2007; Nuovo Cim.B122:1165-1170,2007, arXiv:0712.4152v1 [astro-ph]
\bibitem{feasible3} Claudia Quercellini, Luca Amendola, Amedeo Balbi, Paolo Cabella, Miguel Quartin, Real-time Cosmology, Physics Reports, Volume 521, Issue 3, p. 95-134 (2012), arXiv:1011.2646v2 [astro-ph.CO]
\bibitem{feasible4} Pier-Stefano Corasaniti, Dragan Huterer, Alessandro Melchiorri, Exploring the Dark Energy Redshift Desert with the
Sandage-Loeb Test, Phys.Rev.D75:062001,2007, arXiv:astro-ph/0701433v1
\bibitem{feasible5} Abraham Loeb, Direct Measurement of Cosmological Parameters from the Cosmic Deceleration of Extragalactic Objects, Astrophys.J.499:L111-L114,1998, arXiv:astro-ph/9802122v1

\bibitem{dz_hom1} A. Balbi, C. Quercellini, The time evolution of cosmological redshift as a test of dark energy, Mon. Not. R. Astron. Soc. 382, 1623–1629 (2007), arXiv:0704.2350v3 [astro-ph]
\bibitem{dz_hom2} Ruth Lazkoz, Iker Leanizbarrutia, Vincenzo Salzano, Forecast and analysis of the cosmological redshift drift, Eur. Phys. J. C 78, 11 (2018), arXiv:1712.07555v1 [astro-ph.CO] 
\bibitem{dz_hom3} Francisco S. N. Lobo, Jose Pedro Mimoso, Matt Visser, Cosmographic analysis of redshift drift, JCAP 04 (2020) 043, arXiv:2001.11964v3 [gr-qc]

\bibitem{dz_FR} F.A. Teppa Pannia, S.E. Perez Bergliaffa, N. Manske, Cosmography and the redshift drift in Palatini f(R) theories, Eur. Phys. J. C (2019) 79: 267, arXiv:1811.08176v2 [gr-qc]

\bibitem{dz_varying_c} Adam Balcerzak, Mariusz P. Dabrowski, Redshift drift in varying speed of light cosmology, Phys. Lett. B728, 15-18 (2014), arXiv:1310.7231v2 [astro-ph.CO]

\bibitem{dz_LTB1} C.-M. Yoo, T. Kai, and K.-I. Nakao, Redshift drift in Lemaıtre-Tolman-Bondi void universes, Phys. Rev. D 83, 043527 (2011), arXiv:1010.0091 [astro-ph.CO]
\bibitem{dz_LTB2} S. M. Koksbang and S. Hannestad, Redshift drift in an inhomogeneous universe: averaging and the backreaction conjecture, JCAP 01, 009, arXiv:1512.05624 [astroph.CO]
\bibitem{dz_LTB3} Sofie Marie Koksbang, Asta Heinesen, Redshift drift in a universe with structure I: Lemaitre-Tolman-Bondi structures with arbitrary angle of entry of light, Phys. Rev. D 106, 043501 (2022), arXiv:2205.11907v2 [astro-ph.CO]
\bibitem{dz_LTB4} R. Codur, C. Marinoni, Redshift drift in radially inhomogeneous Lemaitre-Tolman-Bondi spacetimes, Phys. Rev. D 104, 123531 (2021),  arXiv:2107.04868v2 [gr-qc]
\bibitem{dz_SZ1} P. Mishra, M.-N. Celerier, and T. P. Singh, Redshift drift in axially symmetric quasi-spherical Szekeres models, Phys. Rev. D 86, 083520 (2012), arXiv:1206.6026 [astro-ph.CO]
\bibitem{dz_Sz2} P. Mishra and M.-N. Celerier, Redshift and redshift drift in $\Lambda = 0$ quasispherical Szekeres cosmological models and the effect of averaging, Phys. Rev. D 105, 063520 (2022), arXiv:1403.5229 [astro-ph.CO]
\bibitem{dz_stephani1} A. Balcerzak and M. P. Dabrowski, Redshift drift in a pressure gradient cosmology, Phys. Rev. D 87, 063506 (2013), arXiv:1210.6331 [astro-ph.CO]
\bibitem{dz_stephani2} A. Balcerzak, Redshift drift and inhomogeneities, AIP Conf. Proc. 1514, 128 (2013)
\bibitem{dz_BianchiI_1} P. Fleury, C. Pitrou, and J.-P. Uzan, Light propagation in a homogeneous and anisotropic universe, Phys. Rev. D 91, 043511 (2015), arXiv:1410.8473[gr-qc]
\bibitem{dz_BianchiI_2} S. M. Koksbang, Observations in statistically homogeneous, locally inhomogeneous cosmological toy-models without FLRW backgrounds, Mon. Not. Roy. Astron. Soc. 498, L135 (2020), [Erratum: Mon.Not.Roy.Astron.Soc. 500, (2021)], arXiv:2008.07108 [astro-ph.CO]
\bibitem{dz_pert_bianci} Oton H. Marcori, Cyril Pitrou, Jean-Philippe Uzan, Thiago S. Pereira, Direction and redshift drifts for general observers and their applications in cosmology, Phys. Rev. D 98, 023517 (2018), arXiv:1805.12121v1 [astro-ph.CO]
\bibitem{Linder_dz1} Alex G. Kim, Eric V. Linder, Jerry Edelstein, David Erskine, Giving Cosmic Redshift Drift a Whirl, Astroparticle Physics 62, 195 (2015), arXiv:1402.6614v2 [astro-ph.CO]
\bibitem{Linder_dz2} Eric V. Linder, Constraining Models of Dark Energy, arXiv:1004.4646v1 [astro-ph.CO]

\bibitem{dz_ES} S. M. Koksbang, On the relationship between mean observations, spatial averages and the Dyer-Roeder approximation in Einstein-Straus models, JCAP11(2020)061, arXiv:2010.04500v2 [astro-ph.CO]  




\bibitem{dz_test1} J.-P. Uzan, C. Clarkson, and G. F. R. Ellis, Time drift of cosmological redshifts as a test of the Copernican principle, Phys. Rev. Lett. 100, 191303 (2008), arXiv:0801.0068 [astro-ph]
\bibitem{dz_test2} S. M. Koksbang, Quantifying effects of inhomogeneities and curvature on gravitational wave standard siren measurements of H(z), 	Phys. Rev. D 106, 063514 (2022), arXiv:2208.12450v2 [astro-ph.CO]
\bibitem{dz_test3} S. M. Koksbang, Searching for signals of inhomogeneity using multiple probes of the cosmic expansion rate H(z), Phys. Rev. Lett. 126, 231101 (2021), arXiv:2105.11880v1 [astro-ph.CO]

\bibitem{another_look} S. M. Koksbang, Another look at redshift drift and the backreaction conjecture, JCAP10(2019)036, arXiv:1909.13489v1 [astro-ph.CO]



\bibitem{fluid1}  Thomas Buchert: On average properties of inhomogeneous fluids in general relativity I: dust cosmologies, Gen.Rel.Grav. 32 (2000) 105-125, arXiv:gr-qc/9906015v2
\bibitem{fluid2} Thomas Buchert, On average properties of inhomogeneous fluids in general relativity II: perfect fluid cosmologies, Gen.Rel.Grav.33:1381-1405,2001, arXiv:grqc/0102049v2
\bibitem{fluid3} Thomas Buchert, Pierre Mourier, Xavier Roy, On average properties of inhomogeneous fluids in general relativity III: general fluid cosmologies, Gen. Rel. Grav. 52 (2020) 27, arXiv:1912.04213v2 [gr-qc]


\bibitem{Asta_dz_DE} Asta Heinesen, Redshift drift as a model independent probe of dark energy, Phys. Rev. D 103, L081302 (2021), arXiv:2102.03774v1 [gr-qc] 

\bibitem{timescape} David L. Wiltshire, Average observational quantities in the timescape cosmology, Phys.Rev.D80:123512,2009, arXiv:0909.0749v2 [astro-ph.CO] 
\bibitem{timescape2} David L. Wiltshire, Gravitational energy as dark energy: Average observational quantities, AIP Conf.Proc.1241:1182-1191,2010, arXiv:0912.5236v1 [astro-ph.CO]


\bibitem{pec_acc_not1} K. Lake, Remark on Peculiar Corrections to the Time Evolution of the Cosmological Redshift, Astrophysical Letters, Vol. 23, p.23
\\
See e.g. also S. Phillipps, On the Time Evolution of the Cosmological Redshift, Astrophysical Letters, Vol. 22, p.123, 1982
\bibitem{pec_acc_not2} Madhura Killedar, Geraint F. Lewis, Lyman alpha absorbers in motion: consequences of gravitational lensing for the cosmological redshift drift experiment, Mon. Not. R. Astron. Soc. 402, 650–656 (2010), arXiv:0910.4580v1 [astro-ph.CO]
\bibitem{pec_acc_not3} Ryan Cooke, The ACCELERATION programme: I. Cosmology with the redshift drift, Mon. Not. R. Astron. Soc., 492, 2044–2057 (2020), arXiv:1912.04983v1 [astro-ph.CO]


\bibitem{Asta_first_dz} Asta Heinesen, Multipole decomposition of redshift drift -- model independent mapping of the expansion history of the Universe, Phys. Rev. D 103, 023537 (2021), arXiv:2011.10048v1 [gr-qc]


\bibitem{optical_effects} Mikolaj Korzynski, Jaroslaw Kopinski, Optical drift effects in general relativity, JCAP03(2018)012, arXiv:1711.00584v4 [gr-qc]
\bibitem{bigOnLight}  M. Grasso and E. Villa, BiGONLight: light propagation with bilocal operators in numerical relativity, Class. Quant. Grav. 39, 015011 (2022), arXiv:2107.06306 [grqc]

\bibitem{Asta_cosmography} Asta Heinesen, Redshift drift cosmography for model-independent cosmological inference, Phys. Rev. D 104 (2021), 123527, arXiv:2107.08674v2 [astro-ph.CO]
\bibitem{nonlinearities} M. Grasso, E. Villa, M. Korzynski, and S. Matarrese, Isolating nonlinearities of light propagation in inhomogeneous cosmologies, Phys. Rev. D 104, 043508 (2021),arXiv:2105.04552 [astro-ph.CO]



\bibitem{parallax} Syksy Rasanen, A covariant treatment of cosmic parallax, JCAP03(2014)035, arXiv:1312.5738v3 [astro-ph.CO]
\bibitem{position_drift1} Andrzej Krasinski, Spacetimes with no position drift, arXiv:2208.10440v1 [gr-qc]
\bibitem{position_drift2} Andrzej Krasinski, Krzysztof Bolejko, Exact inhomogeneous models and the drift of light rays induced by nonsymmetric flow of the cosmic medium, International Journal of Modern Physics D Vol. 22, No. 06, 1330013 (2013), arXiv:1212.4697v3 [gr-qc]
\bibitem{position_drift3} Andrzej Krasinski, Repeatable light paths in the conformally flat cosmological models, 	Phys. Rev. D86, 064001 (2012), arXiv:1205.6341v3 [gr-qc] 

\bibitem{LTB1} G. Lemaitre: L'Universe en expansion, Annales de la Societe Scientifique de Bruxelles A 53, 51 (1933), English translation: The expanding universe, Gen. Rel. Grav. 29, 637 (1997)
\bibitem{LTB2} R. C. Tolman: Effect of Inhomogeneity on Cosmological Models, Proc. Natl. Acad. Sci. USA 20, 169-176 (1934)
\bibitem{LTB3} H. Bondi: Spherically Symmetrical Models in General Relativity, Month. Not. Roy. Astr. Soc. 107,410 (1947)


\bibitem{Gadget} Volker Springel, Naoki Yoshida, Simon D.M. White, GADGET: A code for collisionless and gasdynamical cosmological simulations, New Astron.6:79,2001, arXiv:astro-ph/0003162v3

\bibitem{Gadget2} Volker Springel, The cosmological simulation code GADGET-2, 	Mon.Not.Roy.Astron.Soc. 364 (2005) 1105-1134, arXiv:astro-ph/0505010v1

\bibitem{GreenWaldMap}  S. R. Green and R. M. Wald, Newtonian and Relativistic Cosmologies, Phys. Rev. D85 (2012) 063512, arXvi:1111.2997


\bibitem{syksy_light} Syksy Rasanen, Light propagation in statistically homogeneous and isotropic universes with general matter content, JCAP 1003:018,2010, arXiv:0912.3370v2 [astro-ph.CO]

\bibitem{acc_dz} Luca Amendola, Claudia Quercellini, Amedeo Balbi, Peculiar acceleration, Phys.Lett.B660:81-86,2008, arXiv:0708.1132v1 [astro-ph]
\bibitem{variance_dz} Jean-Philippe Uzan, Francis Bernardeau, Yannick Mellier, Time drift of cosmological redshifts and its variance, Phys.Rev.D77:021301,2008, arXiv:0711.1950v2 [astro-ph]



\bibitem{dr} S. M. Koksbang, Understanding the Dyer-Roeder approximation as a consequence of local cancellations of projected shear and expansion rate fluctuations, Phys. Rev. D 104, 043505 (2021), arXiv:2106.12913 [astro-ph.CO]

\bibitem{syksy_pert} Syksy Rasanen, Light propagation and the average expansion rate in near-FRW universes, Phys. Rev. D 85, 083528 (2012), arXiv:1107.1176v2 [astro-ph.CO]
\bibitem{tardis} Mikko Lavinto, Syksy Rasanen, Sebastian J. Szybka, Average expansion rate and light propagation in a cosmological Tardis spacetime, JCAP12(2013)051, arXiv:1308.6731v2 [astro-ph.CO]
\bibitem{cheese} S. M. Koksbang, Light propagation in Swiss cheese models of random close-packed Szekeres structures: Effects of anisotropy and comparisons with perturbative results, Phys. Rev. D 95, 063532 (2017), arXiv:1703.03572v2 [astro-ph.CO]

\bibitem{kinematical} Thomas Buchert, Dark Energy from structure: a status report, Gen.Rel.Grav.40:467-527,2008, arXiv:0707.2153v3 [gr-qc]
\bibitem{kinematical2} E.W. Kolb, S. Matarrese, A. Riotto, On cosmic acceleration without dark energy, New J.Phys.8:322,2006,  arXiv:astro-ph/0506534v2

\bibitem{lightcone}  M. Gasperini, G. Marozzi, F. Nugier and G. Veneziano, Light-cone averaging in cosmology: Formalism and applications, JCAP 1107, 008 (2011), arXiv:1104.1167 [astro-ph.CO]
\\
G. Fanizza, M. Gasperini, G. Marozzi, G. Veneziano, A new approach to the propagation of light-like signals in perturbed cosmological backgrounds, JCAP08(2015)020, arXiv:1506.02003v2 [astro-ph.CO] 
\\
Ermis Mitsou, Giuseppe Fanizza, Nastassia Grimm, Jaiyul Yoo, Cutting out the cosmological middle man: General Relativity in the light-cone coordinates, Class.Quant.Grav. 38 (2021) no. 5, 055011, arXiv:2009.14687v2 [gr-qc]\\
Giuseppe Fanizza, Giovanni Marozzi, Matheus Medeiros, Gloria Schiaffino, The Cosmological Perturbation Theory on the Geodesic Light-Cone background, JCAP 02 (2021) 014, arXiv:2009.14134v2 [gr-qc]


\bibitem{null_cone} Thomas Buchert, Henk van Elst, Asta Heinesen, The averaging problem on the past null cone in inhomogeneous dust cosmologies,  arXiv:2202.10798v1 [gr-qc]

\bibitem{quiet} Asta Heinesen, Hayley J. Macpherson, A prediction for anisotropies in the nearby Hubble flow, JCAP03(2022)057, arXiv:2111.14423v1 [astro-ph.CO] 

\bibitem{acoleyen} Karel Van Acoleyen, LTB solutions in Newtonian gauge: from strong to weak fields, JCAP0810:028,2008, arXiv:0808.3554v2 [gr-qc]


\bibitem{ishak} Anthony Nwankwo, Mustapha Ishak, John Thompson, Luminosity distance and redshift in the Szekeres inhomogeneous cosmological models, JCAP 1105:028, 2011, arXiv:1005.2989v3 [astro-ph.CO]

\end{thebibliography}
\end{document}